\documentclass[prodmode,acmcsur]{acmsmall}

\usepackage{algorithmic,amsmath,amssymb,cite,color,comment,fancyhdr,float,graphicx,inputenc,latexsym,ltablex,mathptmx,microtype,setspace,subfigure,tikz,times,url}
\usepackage[ruled]{algorithm2e}

\SetAlFnt{\small}
\SetAlCapFnt{\small}
\SetAlCapNameFnt{\small}
\SetAlCapHSkip{0pt}
\IncMargin{-\parindent}

\usetikzlibrary{decorations.pathreplacing}

\newcommand{\mycomment}[1]{}

\makeatletter
\def\section{\@startsection{section}{1}{\z@}{-3.5ex \@plus -1ex \@minus -.2ex}{2.3ex \@plus .2ex}
  {\normalfont\raggedright\Large\bfseries}}
\def\subsection{\@startsection{subsection}{2}{\z@}{-3.25ex\@plus -1ex \@minus -.2ex}{1.5ex \@plus .2ex}
  {\normalfont\raggedright\large\bfseries}}
\def\subsubsection{\@startsection{subsubsection}{3}{\z@}{-3.25ex\@plus -1ex \@minus -.2ex}{1.5ex \@plus .2ex}
  {\normalfont\raggedright\normalsize\bfseries}}
\makeatother

\begin{document}

\title{Literature Survey on Interplay of Topics, Information Diffusion and Connections on Social Networks}
\author{
KUNTAL DEY
\affil{IBM Research India, and, Indian Institute of Technology, Delhi}
SAROJ KAUSHIK
\affil{Indian Institute of Technology, Delhi}
L. VENKATA SUBRAMANIAM
\affil{IBM Research India}
}

\begin{abstract}
Researchers have attempted to model information diffusion and topic trends and lifecycle on online social networks.
They have investigated the role of content, social connections and communities, familiarity and behavioral similarity in this context.
The current article presents a survey of representative models that perform topic analysis, capture information diffusion, and explore the properties of social connections in the context of online social networks.
The article concludes with a set of outlines of open problems and possible directions of future research interest.
This article is intended for researchers to identify the current literature, and explore possibilities to improve the art.
\end{abstract}






\begin{bottomstuff}
This work is a part of the first author's part-time PhD at Indian Institute of Technology, Delhi, while working at IBM Research.

Author's addresses:
K. Dey {and} L. V. Subramaniam, IBM Research India, New Delhi, India;
S. Kaushik, Computer Science and Engineering Department, Indian Institute of Technology, Delhi, India.
\end{bottomstuff}

\maketitle

\section{Introduction}
\label{sec:intro}
The advent of online social networks and media, such as Twitter, Facebook, Instagram and others, has transformed the way that individuals communicate.
In addition, this has also made a radical change to the way that information reaches to individuals.
The social networking platforms acts as a hotbed for user generated content.
Individuals generate content on these platforms, that get passed to other members of the given networks, over user-to-user communication in forms of messages of different lengths and frequencies.
Further, since these platforms have penetrated among billions of people, the scale of user generated content within these networks is unprecedented.
The information content present within these billions of user-generated messages is also unprecedented.

\subsection{Introducing Commonly Used Terms}
\label{sec:commonterms}
A social network is often captured in form of a graph, where the participants are treated as vertices, the boolean existence of communication or friendship between a vertex pair as an edge (which is also often called a {\it link}), the strength of communication/friendship as edge weight (in weighted graphs) and the direction of friendship and/or message flow (such as {\it followership} on Twitter) as edge direction (in directed graphs).
Over the course of research studies addressing social networks, some terms and concepts in analyzing social networks have become well-accepted.
Yet, often in the absence of formal definitions, there are inherent variations to the notions associated with the use (and abuse) of these terms, that have become inseparable ingredients to the literature.
Table~\ref{tab:commonterms} provides an informal introduction to such terms, and the associated intuition, for ease of readers' understanding.

\begin{table}[thb]
\tbl{\label{tab:commonterms}Well-accepted terms and notions in the social network research community}
{
\begin{tabular}{|p{1.5cm}|p{11.5cm}|}
\hline
\textbf{Term} & \textbf{Notion / Intuition found in literature} \\
\hline
Information cascade & This term corresponds to the information content in social network messages, as found in most of the literature, such as \cite{galuba2010outtweeting}, \cite{yang2010modeling} and many others. Literature also treats it as a group of blog posts hyperlinking to other blog posts \cite{leskovec2007patterns}. \\
\hline
Information diffusion & This term captures the movement of information cascades from one participant / portion of the social network to another. Several models attempt to capture the causes and dynamics of information diffusion content (cascades) in the literature, such as \cite{bakshy2012role}, \cite{kwak2010twitter}, \cite{myers2012information} and many others. \\
\hline
Social influence & This term is often used to capture the notion of a latter individual participant of a social network taking an action that is similar to another former participant's action, by way of the latter explicitly or implicitly imitating the action of the former \cite{anagnostopoulos2008influence}. An example of imitation is retweeting on Twitter. Many works in literature model information diffusion taking social influence into consideration, such as \cite{galuba2010outtweeting}, \cite{yang2010modeling}, \cite{goyal2010learning} and others. \\
\hline
Homophily & Familiarity is perceived when two or more individuals know each other (or, in the context of online social networks, befriend with each other or connect to each other). Similarity is perceived when two or more of individuals like one or more shared objects, items, topics {\it etc}. Homophily is the phenomenon of similar people also becoming socially familiar \cite{mcpherson2001birds}. \\
\hline
Social communities & This represent a group of individuals with a large degree of familiarity. The familiarity either follows a certain structure ensuring a notional sufficiently of connections such as maximal cliques \cite{modani2008large}, k-cores \cite{seidman1983network}, k-plexes \cite{seidman1978graph} {\it etc.}, or properties such as high modularity where the connection density within the given group is significantly higher compared to the other individuals belonging to the same social network \cite{newman2006modularity}. \\
\hline
Topic & In general, a topic captures a coherent of set concepts that are semantically/conceptually related to each other. In the context of social network content analysis, a topic notionally corresponds to a set of correlated user-generated concept. In literature, topics are often identified using techniques such as (a) hashtags of microblogs like Twitter (ex: \cite{cunha2011analyzing}), (b) bursty keyword identification (ex: \cite{cataldi2010emerging} and \cite{mathioudakis2010twittermonitor}), and (c) probability distributions of latent concepts over keywords in user generated content (ex: \cite{lau2012line}). \\
\hline
(Geo-social) Spread of topics & This is usually a term used to portray the maximum (or characteristic) geographical span that a topic has reached out, or expected to reach out to. Literature that addresses geo-social spread of topics include \cite{ardon2013spatio}, \cite{nagar2013topical}, \cite{singh2010situation} and many others. \\
\hline
Topic lifecycle & This term notionally corresponds to the temporal span that a topic stays alive from being introduced into the social network, reach its peak of geographical spread and social depth, and decline till the point it no longer exists in the network. Several works analyze topic lifecycle, such as \cite{lau2012line}, \cite{mathioudakis2010twittermonitor}, \cite{ifrim2014event}, \cite{cataldi2010emerging} and many more. \\
\hline
Topical information diffusion & A body of research tends to model information diffusion, seeding from the topics underlying within the information cascade content, such as \cite{ardon2013spatio}, \cite{narang2013discovery}, \cite{nagar2013topical} and others. These works tend to have the topical nature of information diffusion at the heart of their models. \\
\hline
\end{tabular}
}
\end{table}

\subsection{Motivation}
\label{sec:motivation}
Multiple research studies have shown that information diffuses fast over online social networks.
In a pioneering study, \cite{kwak2010twitter} showed that, characteristics of diffusion of information on social microblogging platforms, like Twitter, is similar to news media.
They stimulated the notion that, Twitter-like microblogging networks are hybrid in nature, combining the characteristics of social and information networks, unlike traditional social networks.
In the meantime, another body of research emerged, that attempted to identify topics and spot trending topics being discussed on the online social media.
\cite{mathioudakis2010twittermonitor} designed TwitterMonitor, for detecting and analyzing trends, and studying trend lifecycle.
Using a two-stage approach comprised of detecting and clustering new content generated by users, founded on dictionary learning to detect emerging topics on Twitter, \cite{kasiviswanathan2011emerging} applied their system on streaming data to empirically demonstrate the effectiveness of their approach.
\cite{lu2012trend} attempted to predict topics that would draw attention in future.
Other studies have also been conducted for trend and topic lifecycle analysis on social networks, specifically Twitter, such as \cite{ifrim2014event}, \cite{lau2012line}, \cite{naaman2011hip}, \cite{osborne2012bieber} and \cite{petrovic2010streaming}.

Predicting the existence of social connections between given pairs of individual members of social networks, in form of social links, has been an area of long-standing research.
Link prediction algorithms that use graph properties have existed for long.
Some well-known link prediction methods are the Adamic-Adar method \cite{adamic2003friends}, Jaccard's coefficient \cite{mcgillintroduction}, rooted PageRank \cite{liben2007link}, Katz method \cite{katz1953new} and SimRank \cite{jeh2002simrank}.
\cite{puniyani2010social} investigated the effectiveness of content in social network link prediction, and experimented on Twitter.
\cite{quercia2012tweetlda} proposed a ``supervised topic classification and link prediction system on Twitter''.
Identifying structural communities that form implicitly based upon familiarity within social networks, rather than by explicit interest-based group memberships, has been another area of long-standing research.
There are multiple definitions of communities; however, the modularity method by \cite{newman2006modularity} is arguably the most well-known and well-accepted definition.
Approximation algorithms to compute modularity fast exist, one of the most well-known algorithms being BGLL proposed by \cite{blondel2008fast}.
While links and communities are rooted to the notion of familiarity, another popular topic of research in online social networks is homophily \cite{mcpherson2001birds}.
Homophily is the phenomenon of similar people also being socially familiar.
Studies such as \cite{de2010birds} considered similarity and social familiarity together, to investigate how information diffusion is impacted by homophily.

Understanding social influence, and analyzing its impact on diffusion characteristics in the context of topics and information, such as spread and longevity, has received immense research focus.
Several works have investigated online social networks and microblogs, and have created information diffusion models that account for the effect of influence of the participants.
\cite{goyal2010learning} created an influence model using the Flickr social network graph and user action logs.
Identifying who influences whom and exploring whether participants would propagate the same information in absence of social signals, \cite{bakshy2012role} measured the effect of social networking mediums in information dissemination, and validated on $253$ million subjects.
\cite{yang2010modeling} modeled the ``global influence'' of social network participants, using the rate of information diffusion via the social network.
Many other works have explored influence and its impact on social networks, along the aspects of information diffusion, topics, interest and the lifecycle of topics.

Addressing the geo-temporal aspects of information diffusion on social networks, researchers have attempted to model the evolution that happens to information and topics over time, and across geographical boundaries.
\cite{ardon2013spatio} characterized the diffusion of ideas on social networks by conducting a spatio-temporal analysis.
They showed that popular topics tend to cross regional boundaries aggressively.
\cite{nagar2013topical} found temporal evolution of topical discussions on Twitter to localize geographically, and evolve more strongly at finer geo-spatial granularities.
For instance, they found that, city level discussions evolve more compared to country level.
\cite{achrekar2011predicting} used Twitter data to collect data pertaining to influenza-like diseases.
Using Twitter data, their model could substantially improve the influenza epidemic predictions made from Government's disease control (CDC) data.
Overall, identifying and characterizing topics and information diffusion has received significant research attention.

Clearly, significant research attention has been invested towards modeling information diffusion, correlating the phenomenon with network structures, and investigating the roles and impacts of topics, the lifecycle of topics, influence, familiarity, similarity, homophily and spatio-temporal factors.
In the current article, we conduct a survey of literature that has created significant impact in this space, and explore the details of some of the models and methods that have been widely adopted by researchers.
The aim is to provide an overview of the representative state-of-the-art models, that perform topic analysis, capture information diffusion, and explore the properties of social connections in this context, for online social networks.
We believe our article will be useful for researchers to identify the current literature, and help in identifying what can be improved over the state of the art.

The rest of the paper is organized as follows.
In Section~\ref{sec:topicalcommunities}, we explore the literature for topic based link prediction and community discovery on social networks.
This is followed by a literature survey for information diffusion, and role of user influence, in Section~\ref{sec:topicalflowdiffusion}.
Section~\ref{sec:topiclifecycle} covers the literature addressing lifecycle of topics, covering the inception, spread and evolution of the topics.
The literature addressing the impact of social familiarity and topical (and interest) similarity is covered in Section~\ref{sec:famsim}.
The literature for spatio-temporal analysis of social network discussion topics has been surveyed in Section~\ref{sec:spatiotemporal}.
A high-level discussion of problems of potential interest, and problems where we believe existing solutions can be improved, is provided in Section~\ref{sec:discussions}.
\section{Topic-Based Link Prediction and Community Discovery}
\label{sec:topicalcommunities}

Link prediction is the problem of predicting the existence of social links amongst social network participant pairs.
In traditional literature, the prediction of links has mostly been carried out by investigating social network graph properties.
Since information spreads on online social networks over topics of discussions, predicting links based upon information content essentially gives an intuition of the pathway that given content (information) would diffuse.
This also holds for communities formed on social network graphs, over links inferred from user-generated topical text content.

\begin{table}[thb]
\tbl{\label{tab:topicalcommunities}Literature for content-based link prediction and community identification}
{
\begin{tabular}{|p{2.5cm}|p{5.5cm}|p{4.5cm}|}
\hline
\textbf{Reference} & \textbf{Key Features / Method Overview} & \textbf{Research Outcome} \\
\hline
\cite{puniyani2010social} & Predicts links based upon user-generated content using LDA. & Shows that their content based link prediction outperforms graph-structure based link prediction. \\
\hline
\cite{quercia2012tweetlda} & Creates user profiles from user-generated tweets. Assigns topics to user profiles. Measures similarity of user profile pairs using L-LDA and SVM. & Shows that L-LDA outperforms SVM for Twitter user profile classification. Uses profile pair similarity thus obtained as a predictor of social links. \\
\hline
\cite{correa2012itop} & Discovers topical communities on user-generated messages on Twitter. Mines retweets, replies and mentions as user-generated indicative signals. Infers global topic-specific communities. & Shows the effectiveness of their method by evaluating communities across three dimensions, namely graph (friendship connections), empirical (actual user profiles) and semantic (frequent n-grams). \\
\hline
\end{tabular}
}
\end{table}

\subsection{Topic-Based Link Prediction}
\label{sec:linkprediction}

Several works in literature, such as \cite{yin2011structural} and \cite{dong2012link}, have addressed predicting social links between pairs of users, looking at the graph attributes.
\cite{liben2007link} carried out a detailed exploration of different link prediction techniques in a social network setting, including methods such as graph distance, Adamic-Adar method \cite{adamic2003friends}, Jaccard's coefficient \cite{mcgillintroduction}, rooted Pagerank \cite{liben2007link}, Katz method \cite{katz1953new} and SimRank \cite{jeh2002simrank}.
Language-based conversation modeling of Twitter users have been carried out by works such as \cite{ritter2010unsupervised}.
However, these studies explore graph structure and properties, and do not consider content semantics.

One body of work uses user-generated as the foundation of the link prediction process.
In one such work, \cite{puniyani2010social} study the effectiveness of content in predicting links on social networks, using Twitter data for experiments.
Using Twitter's {\it GardenHose} API, they collect around $15\%$ of all messages on Twitter, posted in January 2010.
The extract a representative subset by sampling the first $500$ people who posted at least $16$ messages within this period, and subsequently crawl $500$ randomly selected followers of each of these people.
They end up with a data set comprising of $21,306$ users, $837,879$ messages, and $10,578,934$ word tokens posted as part of these messages.

Subsequently, they tokenize while factoring for the non-standard orthography that is inherent to Twitter messages.
They tokenize on whitespaces and apostrophes.
They use the \# mark to indicate a topic, and the @ mark to indicated retweets.
Removing the low-frequency words that appear less than $50$ times from the vocabulary, they are left with $11,425$ tokens.
They classify out-of-vocabulary items were classified as either words, URLs, or numbers.

They use LDA \cite{blei2003latent} for predicting pairwise links on the content graph.
To do so, they gather together all of the messages from a given user into a single document, as the length of Twitter messages are short.
Thus, their model learns latent topics that characterize authors, rather than messages.
They subsequently compute author similarity using dot product of topic proportions.
They learn weight proportions of each topic $z$ using the method of Chang and Blei \cite{chang2010hierarchical} as $ exp(-\eta^T(\bar{z_i} - \bar{z_j})\circ(\bar{z_i} - \bar{z_j}) - \nu)$, as the predicted strength of connection between authors {\it i} and {\it j}.
$\bar{z_i}$ and $\bar{z_j}$ denote the expected topic proportions for author {\it i} and {\it j}, $\eta$ denotes a vector of learned regression weights, and $\nu$ is an intercept term necessary if a the link prediction function returns a probability.
They compare their results with the results obtained by the methodology of Liben-Nowell and Kleinberg \cite{liben2007link}, which depends upon the graph structure but not upon user-generated content.
The content-based model performs significantly better than the structure-based one, establishing a logical foundation to consider user-generated content as an effective instrument to predict social links.

In another work, \cite{quercia2012tweetlda} propose a ``supervised topic classification and link prediction system on Twitter''.
They create user profiles based upon the posts made the by the users.
Their work uses the {\it Labeled-LDA} (L-LDA) technique by \cite{ramage2009labeled}, a generative model for multiply labeled corpora that generates a labeled document collection.
L-LDA assigns one topic to each label in a multiply-labeled document, unlike traditional LDA and its supervised embodiments.
It incorporates supervision to extend LDA \cite{blei2003latent} and incorporates a mixture model to extend Multinomial Naive Bayes.
Each document is modeled as a mix of elemental topics by L-LDA.
Each word is generated from a topic.
The topic model is constrained to only use topics corresponding to a document's observed set of labels.
They ``set the number of topics in L-LDA as the number of unique labels {\it K} in the corpus'', and run LDA such that the multinomial mixture distribution $\theta^{(d)}$ is defined only for topics corresponding to the labels $\Lambda^{(d)}$, the binary list of topics indicating the presence/absence of a topic $l$ inside document $d$.
To enable this constraint, they first generate the document labels $\Lambda^{(d)}$ for each topic $k$ using a Bernoulli coin toss, with a labeling prior probability $\Phi_k$.
They subsequently define the document label vector as: $\lambda^{(d)} = \{k|\Lambda_k^{(d)} = 1\}$.
For each document $d$, a ``document-specific label projection matrix'' $L^{(d)}$ of size $M_d \times K$ is defined by setting for each row $i \in \{1,...,M_d\}$ and for each column $j \in \{1..K\}$ as
\[
	L_{ij}^{(d)} = 
	\begin{cases}
		1 & \text{if } \lambda_i^{(d)} = j \\
		0	& \text{otherwise}
	\end{cases}
\]

If the $i^{th}$ document label and the $j^{th}$ topic are the same, then the $(i,j)^{th}$ element of the $L^{(d)}$ matrix has a value of $1$, else zero.
The ``parameter vector of the Dirichlet prior $\alpha = (\alpha_1,...,\alpha_K)^T$'' uses the $L^{(d)}$ matrix to project to a vector $\alpha^{(d)}$ of a lower dimension as
$$ \alpha^{(d)} = L^{(d)} \times \alpha = (\alpha_{\lambda_1^{(d)}},...,\alpha_{\lambda_{M_d}^{(d)}})^T $$
The dimensions of the $\alpha^{(d)}$ vector ``correspond to the topics represented by the document labels''.
Finally, $\theta^{(d)}$ is drawn from this Dirichlet distribution.

They experiment on Twitter data using the L-LDA technique.
They assign topics to user profiles, and measured the similarity of user profile pairs.
They find L-LDA to significantly outperform Support Vector Machines (SVM) for user profile classification, in cases where the training data is limited, and provide similar performance as SVM where sufficient training data is available.
They thereby infer L-LDA to be a good technique to classify infrequent topics and (short) profiles of users having moderate activity.
They treat user profile pair similarities as predictor of social links.

\subsection{Topic-Based Community Discovery}
\label{sec:communityprediction}

In the social network analysis literature, communities are identified by one of the following.
(a) Individuals subscribe to existing interest groups, and thereby start explicitly belonging to a community based upon their similarity of interests.
(b) Groups of individuals known to each other directly, or having a large number of mutual friends, are said to belong to the same implicit community.
While several definitions of structural communities have emerged over time, modularity-based community finding \cite{newman2006modularity} is the most popular methodology.
Modularity-based community finding from a given graph is inherently expensive.
\cite{blondel2008fast} propose BGLL as a fast approximation algorithm towards this.
\cite{yang2015defining} investigate structural and functional communities, and impacts of structure on community functions.

Literature mostly explores community discovery from explicit links such as social friendships.
However, some work also exists to find communities formed upon links inferred from user-generated topics and/or content.
In one such work, \cite{correa2012itop} discover topical communities on Twitter tweets.
They mine retweets, replies and mentions, collectively labeling these as @-messages.
They create an edge between a vertex (user) pair $v_x$ and $v_y$ if $I(RT_{xy},@_{xy}) \neq 0$, where $I(RT_{xy},@_{xy})$ is the @-message based interaction strength between $v_x$ and $v_y$.
They adapt the local modularity (LM) algorithm \cite{clauset2005finding} for directed graphs, to discover communities of interest using local information.
Their framework comprises of four blocks: {\it warm start}, {\it expand}, {\it filter} and {\it iterate}.

For {\it warm start}, they take a topic of interest $t_i$ as input, and conducts a Twitter user {\it bio} search, where {\it bio} comprises of the publicly available profile information of the user such as name, location, URL and biography.
The list of users found to have related interest and inclination towards this topic, as found by the search, are included as parts of communities of interest, denoted as $C^{t_i}_{current}$.
In the {\it expand} step, they take this list of users, and adds vertices $U^{t_i}$, where $\beta^{t_i} \in C^{t_i}_{current}$ has an edge with at least one vertex in $U^{t_i}$.
The weight of an edge is defined by the closeness of the user pair in terms of @-messages.
For instance, a directed edge $X \rightarrow Y$ is drawn from $X$ to $Y$, iff $X$ has interacted with $Y$.
Further, weight $w$ is assigned based upon the interaction strength.
This, graph $G^{t_i}_{current}$ gets created by the expansion process, where the vertex set $V^{t_i}_{current} = U^{t_i} \cup C^{t_i}_{current}$ and $U^{t_i} \cap C^{t_i}_{current} = \phi$.

In the {\it filter} step, they iterate through each vertex $v_i \in U^{t_i}$ of the previous ({\it expand}) step.
They compute the local modularity $R$ for each vertex $v_i$.
Formally, $R \propto \frac{1}{\beta^{t_i}}$, and measures the sharpness of $\beta^{t_i}$.
$R_{v_i}$ is the ratio of weighted edges with no node in $U^{t_i}$ and with at least one node in $\beta^{t_i}$ for all $v_i$.
The algorithm performs a greedy maximization of local modularity.
The vertex yielding maximum local modularity, $R_{max}^{v_i}$, is added to $C^{t_i}_{current}$.
The {\it iterate} step ensures that the process repeats the {\it expand} and {\it filter} steps, until the stopping condition is reached.
The stopping condition checks whether $R_{max}^{v_i}$ is stable or consistently negative, indicating that there is no further place for improvement.
Thus, they identify topic-specific global communities, taking topic as an input keyword.
They ``evaluate the communities along three dimensions, namely graph (vertex-edge quality), empirical (actual Twitter profiles) and semantic (n-grams frequently appearing in tweets)''.

In another work, \cite{weng2014topic}, explore the Facebook social network for topic based cluster analysis, and shows that friends that favor similar topics form topic-based clusters.
This study further shows that these clusters have dense connectivity, large growth rate, and little overlap.

Cross-entropy \cite{de2005tutorial}, which is based upon Kullback-Leibler (K-L) divergence \cite{kullback1951information}, and normalized mutual information \cite{coombs1970mathematical}, are relevant measurements frequently appearing in literature of communities, user profile pair similarities and topical divergence computation.

\section{Information Diffusion and Role of Influence}
\label{sec:topicalflowdiffusion}

Diffusion of information content on social networks such as Twitter and Facebook, has been a major research focus \cite{fox2011social} \cite{howard2011opening} \cite{hughes2009twitter} \cite{sun2009gesundheit}.
Several information diffusion models, such as Linear Threshold \cite{granovetter1978threshold} and Independent Cascades \cite{goldenberg2001talk}, and variations of these models, have been built.
Models have attempted to capture diffusion path, degree of diffusion for specific information on observed social networks, and the role of influence of participants in the information flow process.

\begin{table}[thb]
\tbl{\label{tab:informationdiffusion}Literature for information diffusion and role of influence}
{
\begin{tabular}{|p{3cm}|p{5.5cm}|p{5cm}|}
\hline
\textbf{Reference} & \textbf{Key Features / Method Overview} & \textbf{Research Outcome} \\
\hline
\cite{kwak2010twitter} & Analyzes topological characteristics, followers versus tweets distribution, reciprocity, degree of separation and homophily. Ranks Twitter users by PageRank and retweets. Compares singletons, replies, mentions and retweets with trends in other media. Conducts temporal analysis. & Shows that information diffuses on Twitter like news media. Shows that irrespective of follower count of the original tweet writer, a tweet reaches to about 1,000 users on an average. Shows that Twitter (a microblog) combines the aspects of traditional social networks and information networks. \\
\hline
\cite{galuba2010outtweeting} & Characterizes the propagation of URLs on the Twitter platform. Shows statistical regularities in user activity, social graph, URL cascade structure and communication dynamics, on $2.7$ million users exchanging $15$ million URLs. & Proposes a propagation model predicting which URL will each given user mention, and shows the effectiveness of their model. \\
\hline
\cite{kitsak2010identification} & Identifies a network core using k-shell decomposition analysis, where the more central vertices in the graph receive higher k-values. The innermost vertices form the graph core. & Shows that the network core members are best spreaders of information, not the most highly connected or the most centrally located ones. \\
\hline
\cite{kossinets2008structure} & Formulates a temporal notion of social network distance measuring the minimum time for information to spread across a given vertex pair. Defines a network backbone, a subgraph in which the information flows the quickest. & Shows that the network backbone for information propagation on a social network graph is sparse, with a mix of ``highly embedded edges and long-range bridges''. \\
\hline
\cite{bakshy2012role} & Quantifies the causal effect of social networks in disseminating information, by identifying who influences whom, and exploring whether they would propagate the same information if the social signals were absent. Experiments with information sharing behavior of $253$ million users. & Shows that while stronger ties are more influential at an individual level, the abundance of weak ties are more responsible for novel information propagation. \\
\hline
\cite{de2010birds} & Hypothesizes that homophily affects the core mechanism behind social information propagation. Proposes a dynamic Bayesian network for capturing information diffusion. & Shows that considering homophily leads to an improvement of $15\%$-$25\%$ in prediction of information diffusion. \\
\hline
\cite{yang2010modeling} & Models the global influence of a node, on the ``rate of information diffusion through the implicit social network''. Proposes Linear Influence Model, in which a newly infected (informed) node is a ``function of other nodes infected in the past''. & Shows that the patterns of influence of individual participants significantly differs, depending on node type and topic of information. \\
\hline
\cite{yang2015defining} & Explores speed, scale and range as major properties of social network information diffusion. & Shows that user properties, and the rate at which a user is mentioned, are predictors of information propagation. Shows that information propagation range for an event is higher for tweets made later. \\
\hline
\cite{myers2012information} & Observes that information can flow both through online social networks and sources outside the network such as news media. Models information propagation accordingly. Uses hazard functions to quantify external exposure and influence. Applies the model to URLs emerging on Twitter. & Shows that, affected by external influence (and not social edges), information jumps across the Twitter network. Quantifies information jump. Shows that $71\%$ of information diffuses over Twitter network, while $29\%$ happens out of the network. \\
\hline
\end{tabular}
}
\end{table}

\subsection{Topical Information Diffusion on Social Networks}
\label{subsec:information}

In a pioneering study, \cite{kwak2010twitter} suggest information diffuses on Twitter-like social microblogging platforms in a similar manner as news media.
They show that, over the original tweet and retweets, and regardless of the followers of the originator of the tweet, a tweet reaches to about 1,000 users on an average.
It stimulates the notion that, such microblogging networks are hybrid in nature, where the characteristics of social and information networks get combined.
Their dataset comprises $41.7$ million Twitter users, $1.47$ billion social followership edges and $106$ million tweets.
They observe that Twitter trends are different from traditional social network trends, with lower than expected degrees of separation, and non-power-law distribution of followers.
The reciprocity of Twitter is low, compared to traditional social networks.
However, the reciprocated relationships exhibit homophily \cite{mcpherson2001birds} to an extent.

They rank Twitter users by PageRank of followings, number of followers and retweets.
They find that the rankings by PageRank and by number of followers are similar, but ranking by retweets is significantly different.
They measure this by using an optimistic approach of the generalization of Kendall's tau \cite{kendall1938new} proposed by \cite{fagin2003comparing}, setting penalty $p = 0$.
They observe that a significant proportion of live news that is of broadcasting nature (such as accidents and sports), breaks out on Twitter ahead of CNN, a traditional online media.
They note that around $20\%$ of Twitter users participate in trending topics, and around $15\%$ of the participants participate in more than $10$ topics in $4$ months.
They observe that the active periods of most trends are a week or shorter.

They attempt to investigate whether favoritism exists in retweets.
For this, assuming user $j$ makes $|r_{ij}|$ retweets to user $i$, they define $Y(k,i)$ as
$$ Y(k,i) = \sum\limits_{j=1}^k\Bigg\lbrace\frac{|r_{ij}|}{\sum\limits_{l=1}^k|r_{il}|}\Bigg\rbrace^2 $$
$Y(k)$ averages $Y(k,i)$ over all vertices having made / received $k$ retweets.
If followers tend to evenly retweet, then $kY(k) \sim 1$.
And $kY(k) \sim k$ if only a subset of followers retweet.
Experimentally, they observe linear correlation to $k$, which indicates retweets to contain favoritism: people retweet only from a small number of people and only a subset of followers of a user tend to retweet.
In a way, this indicates only a few users to influence the information to diffuse further via retweets, given the user originating the information with respect to the persons retweeting.

\cite{kitsak2010identification} show that the most central or highly connected people are not necessarily the best spreaders of information; often, those located at the network core are.
They identify the best spreaders by {\it k-shell decomposition analysis} \cite{bollobas1984graph} \cite{carmi2007model} \cite{seidman1983network}.
They further show that, when more than one spreader are considered together, the distance between them plays a critical role in determining the spread level.
They apply the Susceptible-Infectious-Recovered (SIR) and Susceptible-Infectious-Susceptible (SIS) models \cite{anderson1992infectious} \cite{heesterbeek2000mathematical} \cite{hethcote2000mathematics} on four different social networks including an email network in a department of a university in London, a blogging community (LiveJournal.com), a contact network of inpatients in a Swedish hospital and ``a network of actors that have co-starred in movies labeled by imdb.com as adult''.
They use a small value of $\beta$, ``the probability that an infectious vertex will infect a susceptible neighbor'', keeping the infected population fraction small.
Using k-shell (k-core) decomposition, they assign coreness $k_S$, an integer index (coreness index), to each vertex of degree $k$, that captures the depth (layer/k-shell) in the network that the vertex belongs to.
The coreness index $k_S$ is assigned such that the more centrally the vertex is located in the graph, the higher is its $k_S$ value.
The innermost vertices thereby form the graph core.

If $(k_S,k)$ is the coreness and degree of vertex $i$ (origin of the epidemic) and ``$\gamma(k_S,k)$ the union of all the $N(k_S,k)$ such vertices'', then the average population $M_i$ infected with the epidemic under SIR-based spreading, averaged over all such origins, is
$$ M(k_S, k) = \sum\limits_{i \in \gamma(k_S,k)}\frac{M_i}{N(k_S,k)} $$
Their analysis finds three general results.
(a) A number of poor spreaders exist among the hubs on the network periphery (large $k$, low $k_S$).
(b) Infected nodes belonging to the same k-shell give rise to similar outbreaks of epidemic, irrespective of the degree of the origin of infection.
(c) The ``inner core of the network'' comprises of the most efficient disease (information spreaders), independent of their degree.
They empirically observe that the influence spreading behavior is better predicted by the k-shell index of a node, compared to the entire network considered as a whole, as well as compared to betweenness centrality.
An outbreak starting at the network core (large $k_S$) finds many paths for the information to spread over the whole network, regardless of the degree of the vertex.
In a subsequent work, \cite{feng2011measuring} modify the k-shell decomposition analysis algorithm to use log-scale mapping, that produces fewer but more appropriate k-shell values.

\cite{kossinets2008structure} propose a temporal notion of social network distance, using the shortest time needed for information to reach to vertex from another.
They find that, structural information that is not evident from analyzing the topology of the social network, can be obtained from such temporal measures.
They define a network backbone, a subgraph in which the information flows the quickest, and experimentally show that the network backbone for information propagation on a social network graph is sparse, with a mix of long-range bridges and strongly embedded edges.
They demonstrate on two email datasets and user communications across Wikipedia admins and editors.

To find the temporal notion of social network distance, they attempt to quantify how updated is each vertex $v$ about each different vertex $u$ at time $t$.
For this, they try to determine the largest $t' < t$ such that, information can reach, from vertex $u$ starting at time $t'$, to $v$ at or before time $t$.
The view of $v$ towards $u$ at time $t$ is the largest value of $t'$, denoted by $\phi_{v,t}(u)$.
They define ``{\it information latency} of $u$ with respect to $v$ at time $t$'' as ``how much $v$'s view of $u$ is {\it out-of-date} at time $t$'', quantified as $(t - \phi_{v,t}(u))$.
Thus iterating over all vertices, they take the view of $v$ to all the vertices in the graph at time $t$, and represent it as a single vector $\phi_{v,t} = (\phi_{v,t}(u) : u \in V)$.
They define $\phi_{v,t}$ as the {\it vector clock} of each vertex $v$ at time $t$.
$\phi_{v,t}$ is updated whenever $v$ receives a communication.

They define the instantaneous backbone of a network using the concept of {\it essential edges}.
In the backbone, ``an edge $(v,w)$ is {\it essential} at time $t$ if the value $\phi_{w,t}(v)$ is the result of a vector-clock update directly from $v$, via some communication event $(v,w,t')$, where $t' < t$''.
Intuitively, an edge $(v,w)$ becomes essential if the most updated view of the target ($w$) of the source ($v$) is via a direct communication over the edge, rather than via an indirect path over other edges.
They define the backbone $H_t$ of the graph at time $t$ to have the vertex set $V$, and the edge set from the original graph $G$ essential at time $t$.
Using this, and assuming a perfectly periodic communication pattern of vertex pairs, they develop a notion of aggregate backbone by aggregating the communication over the entire period of observation.
For each edge $(v,w)$ in $H_t$ where $\rho_{v,w} > 0$ ($v$ has sent $w$ at least one message) within time period $[0,T]$, the {\it delay} $\delta_{v,w}$ is defined for the edge as $T/\rho_{v,w}$, which simply approximates the communication from $v$ to $w$ as temporally evenly spaced.
They assign weight $\delta_{v,w}$ to each edge $(v,w)$, obtaining $G^{\delta}$ from $G$.
In this aggregate setting, where communications are spaced evenly, the path where the sum of the delays is minimum is the path over which information would reach the fastest between that pair of vertices.
They define essential edges in the aggregate sense in $G^{delta}$, and define $H^*$, an {\it aggregate backbone}, constituted using only these essential aggregate edges.

They define the {\it range} of an edge $(v,w)$ as the shortest unweighted alternate path from $v$ to $w$ over the social network, if $e$ was deleted.
On a typical social network, the value of this is often observed to be $2$ as most pairs of social connections tend to have common (shared) friends.
They define the {\it embeddedness} of an edge $(v,w)$ is intuitively the fraction of neighbors common to both $v$ and $w$.
Formally, if $N_v$ and $N_w$ respectively denote the neighbor set of $v$ and $w$, then embeddedness of $e$ is defined as $\frac{|N_v \cap N_w|}{|N_v \cup N_w|}$.
Intuitively, endpoints of edges with high embedding have many common neighbors, hence occupy dense clusters.
Experimentally, they find that highly embedded edges are over-represented both in instantaneous and aggregate backbones.
These represent edges with high rates of communications.
Hence, presence of such edges in the backbone leads to fast information diffusion.
They also observe that, increase in node-dependent delays (delays $\epsilon$ introduced at nodes, in addition to the edge delay $\delta_{v,w}$) in leading to denser backbones.
As that happens, the significance of quick indirect paths diminish.
They note that, to influence the potential information flow, a practical method for individuals is to consider varying the communication rates by simple rules.

\cite{de2010birds} study the impact of user homophily on information diffusion on Twitter data.
They hypothesize that, homophily affects the core mechanism behind social information propagation, by structuring the ego-networks and impacting their communication behavior of individuals.
They follow a three-step approach.
First, for the full social graph (baseline) and filtered graphs using attributes such as activity behavior, location {\it etc.}, they extract diffusion characteristics along categories such as user-based (volume, number of seeds), topology-based (reach, spread) and time (rate).
Second, to predict information diffusion for future time slices, they propose a dynamic Bayesian network.
Third, they use the ability of the predicted characteristics of explaining the ground-truth of observed information diffusion, to quantify the impact of homophily.
They empirically find that, the cases where homophily was considered, could explain information diffusion and external trends by a margin of $15\%$-$25\%$ lower distortion than the cases where it was not considered.

They consider a social action set $O = \{O_1, O_2, ...\}$ (such as, posting a tweet) and a set of attributes $A = \{a_k\}$ (location, organization {\it etc.}).
They consider four user attributes: location, information role (generators, mediators, receptors), content creation (those making self-related posts versus informers), and activity behavior (actions performed on the social network over a given time period).
A pair of users are {\it homophilous} if at least one of their attributes matches more than the random expectation of match in the network.
They construct an induced subgraph $G(a_k)$ of $G$ by selecting vertices, where $G(a_k = v)$ where $v$ is the value of attribute $a_k \in A$ in the selected vertices.
Edges in $G$ are selected in $G(a_k)$, where both the endpoint vertices of the edge are included in $G(a_k)$.
The authors define $s_N(\theta)$, a ``{\it diffusion series} on topic $\theta$ over time slices $t_1$ to $t_N$, as a directed acyclic graph'', in which the vertices correspond to a subset of social network users, involved in social action $O_r$ on topic $\theta$ within time $t_1$ and $t_N$.
Vertices are assigned to slots: all vertices associated with time slice $t_m$ $(t_1 < t_m < t_N)$ are assigned slot $l_m$.

They subsequently attempt to characterize diffusion.
They extract diffusion characteristics on $\theta$ at time slice $t_N$, from each diffusion collection $S_N(\theta)$ (defined as $\{s_N(\theta)\}$) and $\{S_{N;a_k}(\theta)\}$, as $d_N(\theta)$ and $\{D_{N;a_k}(\theta)\}$ respectively.
They use eight different measures to quantify diffusion at each given time slice $t_N$: the volume $v_N(\theta)$ with respect to topic $\theta$ (total volume of contagion present in the graph); participation $p_N(\theta)$ that involve in the information diffusion and further trigger other users to diffuse information; dissemination $\delta_N(\theta)$ that act as seed users of the information diffusion due to unobservable external influence; reach $r_N(\theta)$ to which extent a particular topic $\theta$ reaches to users by the fraction of slots; spread as ratio of the maximum count of informed vertices found over all slots in the diffusion collection, to the total user count; cascade instances $c_N(\theta)$ that defines the fraction of slots in $s_N(\theta) \in S_N(\theta)$ in which the number of new users at slot $l_m$ is higher than the previous slot $l_{m-1}$; collection size $\alpha_N(\theta)$ as the proportion of the number of diffusion series to the number of connected components; and rate $\gamma_N(\theta)$ as the speed of information diffusion on $\theta$ in $S_N(\theta)$.

For each diffusion collection $S_N{\theta}$ and $\{S_{N;a_k}(\theta)\}$, they predict at time slice $t_N$, which users have a higher likelihood of repeating a social action taken at time slice $t_{N+1}$.
This gives diffusion collections at $t_{N+1}$ as: $\hat{S}_{N+1}(\theta)$ and $\{\hat{S}_{N;a_k}(\theta)\} \forall a_k \in A$.
They propose a dynamic Bayesian network, and model the likelihood of action $O_i$ at $t_{N+1}$ using environmental features (activity of a given individual and their friends on a topic $\theta$, and the popularity of the topic $\theta$ in the previous time slot $t_N$) represented by $F_{i,N}(\theta)$ and diffusion collection $S_{i,N+1}(\theta)$.
The goal is to estimate the expectation of social actions: $\hat{O}_{i,N+1} = E(O_{i,N+1}|O_{i,N},F_{i,N})$.
Using first order Markov property, they rewrite this as a probability function: $P(O_{i,N+1}|O_{i,N},F_{i,N}) = \sum\limits_{S_{i,N+1}}P(O_{i,N+1}|S_{i,N+1})P(S_{i,N+1}|S_{i,N},F_{i,N})$.

They use the ``Viterbi algorithm on the observation-state transition matrix to determine the most likely sequence at $t_{N+1}$'', thus predicting the observed action (the first term).
They predict the second term, the hidden states, as $P(S_{i,N+1}|S_{i,N},F_{i,N}) \propto P(F_{i,N}|S_{i,N})P(S_{i,N+1}|S_{i,N})$.
They subsequently substitute the probability of emission $P(O_{i,N+1}|S_{i,N+1})$ and $P(S_{i,N+1}|S_{i,N},F_{i,N})$ to estimate the observed action of $u_i$: $\hat{O}_{i,N+1}$.
They repeat this for each user for time slice $t_{N+1}$.
Using $G$ and $G(a_k)$, they ``associate edges between the predicted user set, and the users in each diffusion series for the diffusion collections at $t_N$''.
They thus obtain the diffusion collection $t_{N+1}$, i.e., $\hat{S}_{N+1}(\theta)$ and $\hat{S}_{N+1;a_k}(\theta)$.

They measure the distortion between actual diffusion characteristics with the predicted, at $t_{n+1}$, using: (a) saturation measurement and (b) utility measurement.
Intuitively, to measure the information content that has diffused into the network on topic $\theta$, saturation measurement is used.
Utility measurement, on the other hand, attempts to correlate the prediction with external phenomena such as search and world news trend.
Using cumulative distribution functions (CDF) of diffusion volume, they model search and news trend measurement models using Kolmogorov-Smirnov (KS) statistic, given as $max(|X - Y|)$ for a given diffusion $D(X,Y)$, where $X$ and $Y$ are two vertices of the graph.

\cite{myers2012information} observe that real-world information can spread via two different ways: (a) over social network connections and (b) over external sources outside the network, such as the mainstream media.
They point out that most of the literature assumes that information only passes over the social network connections, which may not be entirely accurate.
They model information propagation, considering that information can reach to individuals along both the possible ways.
They develop a model parameter fitting technique using hazard functions \cite{elandt1980incomplete} \cite{johnson1999survival}, to quantify the level of external exposure and influence.

In their setting, {\it event profile} captures the ``influence of external sources on the network as a function of time''.
With time, nodes receive ``streams of varying intensity of external exposures, governed by event profile $\lambda_{ext}(t)$''.
A node can get infected by each of the exposures, and eventually the node either becomes infected, or the arrival of exposures cease.
Neighbors receive exposures from infected nodes.
They define exposure curve $\eta(x)$, which determines how likely it is for a node to get infected with arrival of each exposure, and set out to find the shape of the curve, as well as infer how many exposures external sources generate over time.
They model internal exposures using an {\it internal hazard function}, as
$$\lambda_{int}(t)dt \equiv P(i \text{ exposes } j \in [t,t+dt)|i \text{ has not exposed } j \text{ yet})$$
Here $i$ and $j$ are neighbors, and ``time $t$ has passed since node $i$ was infected''.
Intuitively, in their setting, $\lambda_{int}$ effectively models the time taken by a node to understand that one of its nodes have become infected.
The ``expected number of internal exposures node $i$ receives by time $t$'' can be derived by summing up these exposures.
They model exposure to unobserved external information sources, with varying intensities over time, as {\it event profile}, as $\lambda_{ext}(t)dt \equiv P(i \text{ receives exposure } j \in [t,t+dt))$.
The above holds ``for any node $i$, where $t$ is the time elapsed since the current contagion had first appeared in the network''.

They ``model the arrival of exposures as a binomial distribution''.
Since users receive both internal and external exposures simultaneously, they use the average of $\lambda_{ext}(t) + \lambda_{int}^{(i)}(t)$ to ``approximate the flux of exposures as constant in time, such that each time interval has an equal probability of arrival of exposures''.
The ``sum of these events is a standard binomial random variable''.
If a node receives $x$ exposures where the exposure curve is $\eta(x)$, then $\eta(x)$ is computed as:
$$ \eta(x) = P(\text{node }i\text{ is infected immediately after }x^{th}\text{ exposure}) \\
= \frac{\rho_1}{\rho_2}. x . exp\bigg(1 - \frac{x}{\rho_2}\bigg)$$

Here $\rho_1 \in (0,1]$ and $\rho_2 > 0$.
Note that, $\eta(0) = 0$.
This implies, a node can be infected only after being exposed to a contagion.
The function is unimodal with an exponential tail.
Hence there a critical mass of exposures exists when the contagion is most infectious.
This is followed by decay, caused by becoming overexposed/tiresome.
Importantly, $\rho_1 = max_x\eta(x)$ measures the infectiousness of a contagion in the network, and $\rho_2 = arg max_x\eta(x)$ measures the contagion's enduring relevancy.

For a given node $i$, the infection time distribution can be built as following.
Let $F^{(i)}(t) \equiv P(\tau_i \leq t)$ denote ``the probability of node $i$ being infected by time $t$'', where node $i$ has been infected at time $\tau_i$.
Using $P_{exp}^{(i)}(n;t)$, $F^{(i)}(t)$ is derived as
$$ F^{(i)}(t) = \sum\limits_{n=1}^{\infty}P_{exp}^{(i)}(n;t) \times \Bigg[1 - \prod\limits_{k=1}^n[1 - \eta(k)] \Bigg] $$
Although $F^{(i)}(t)$ is ``analogous to the cumulative distribution function of infection probability'', it is ``not actually a distribution'': $\lim_{x \rightarrow \infty}\eta(x) = 0$ leads to $\lim_{t \rightarrow \infty}F(t) < 1$.
Their model also ensures that the chance of a node never becoming infected is non-zero, which is realistic.

They apply the model to the URLs emerging on Twitter.
They observe that information {\it jumps} across the Twitter network that the social edges cannot explain, and is necessarily caused by unobservable external influences.
They quantify the information jump, noting that around $71\%$ of information diffuses over the Twitter network, while the other $29\%$ happens over external events outside.

\cite{wu2014opinionflow} create an interactive visualization tool, to visually summarize the opinion diffusion by using a combination of a Sankey \cite{sankey1898introductory} graph and a tailored density map, at a topic level.
Using an information diffusion model that uses a combination of reach (the average number of people influenced by message published by a given user), amplification (the likelihood that audience responds to a message) and network score (the influence of a users' audience) to measure user influence levels, they characterize the propagation of opinions among many users regarding different topics on social media.
\cite{narayanan2009anonymizing} aim to identify (de-anonymize) users across social networking platforms.
They hypothesize that, identifying the profiles of users across multiple social networking platforms would provide more insights into the information diffusion process, by observing the diffusion of information over these multiple platforms at a given time.
They demonstrate their hypothesis using Twitter and Flickr in combination.
\cite{tsur2012s} attempt to predict the spread of ideas on Twitter, combining topological and temporal features with content features, for minimizing errors.
\cite{lerman2010information} empirically study the characteristics of news spreading on several popular social networks, such as Twitter and Digg.
\cite{hong2011predicting} propose a multi-class classification model to identify popular messages on Twitter, by predicting retweet quantities, from TF-IDF (term frequency and inverted document frequency) and LDA, along with social properties of users.

\subsection{The Role of Influence}
\label{subsec:influence}

Social influence plays a significant role in information diffusion dynamics \cite{granovetter1978threshold} \cite{watts1998collective}.
Research has attempted to investigate information cascade flow along underlying social connection graphs, and analyze the role of influence in such propagation.

\cite{cha2010measuring} explore influence on Twitter based on indegree, mentions and retweets.
They find that individuals with high indegree do not necessarily generate many mentions and retweets.
They observe that while majority of influential users tend to influence several topics, influence is gathered through focused efforts, such as limiting tweets to one topic.
\cite{bakshy2011everyone} study influencing behavior in terms of cascade spread on Twitter.
They find that the past influence of users and the interestingness of content can be used to predict the influencers.
They observe that although URLS rated interesting, and content by influential users, spread more than average, no reliable method exists for predicting which user or URL will generate large cascades.
\cite{liu2010mining} study social influence in large scale networks using a topical sum-product algorithm, and investigate the impact of topics in social influence propagation.
\cite{romero2011differences} study the role of passivity and propose a PageRank like measure to find influence on Twitter.
\cite{weng2010twitterrank} too propose a PageRank like measure to quantify influence on Twitter, based on link reciprocity and homophily.
\cite{bi2014scalable} and \cite{cano2014social} conduct topic-specific influence analyses for microblogs.

\cite{galuba2010outtweeting} characterize the propagation of URLs on Twitter, and predict information cascades, factoring for the influence of users on one another.
Tracking $2.7$ million users exchanging over $15$ million URLs, they show statistical regularities to be present in social graph, activity of users, URL cascade structure and communication dynamics.
They look at URL sharing activities such as URL mentions by users in their tweets, URL popularity (how frequently they appears in tweets) and user activity (how frequently they mention URLs).
They define two information cascade types.
In {\it F-cascade}, the flow of URLs are constrained to the follower graph.
They draw an edge between a vertex pair $v1$ and $v2$ iff: (a) ``$v1$ and $v2$ tweeted about URL $u$'', (b) ``$v1$ mentioned $u$ before $v2$'', and (c) ``$v2$ is a follower of $v1$''.
In {\it RT-cascade}, they use a who-credits-whom model.
They disregard the follower graph, and draw an edge between $v1$ and $v2$ iff: (a) ``$v1$ tweeted about URL $u$'', (b) ``$v1$ mentioned $u$ before $v2$'', and (c) ``$v2$ credited $v1$ as the source of $u$''.
Using this, they proposed a propagation model predicting which URLs are likely to be mentioned by which users.

They construct two information diffusion models.
The {\it At-Least-One} (ALO) model assumes it sufficient to cause a user to tweet by influence of one user.
Retweet probability in ALO model is computed as $p_i^u = A(\alpha_{ji},\beta_i,\gamma_u).T(\mu_i,\sigma_i^2,t_i^u)$.
They define the time-independent component $A$ as $A(\alpha_{ji},\beta_i,\gamma_u) = 1 - (1 - \gamma_u\beta_i)\prod\limits_{j:i \rightarrow j}(1 - \gamma_u\alpha_{ji}p_j^u)$.
The $\alpha_{ji} \in [0,1]$ parameters capture the influence of $j$ on $i$ ($j$ follows $i$), $\beta_i \in [0,1]$ is the ``baseline probability of user $i$ tweeting any URL'' and $\gamma_u \in [0,1]$ is the virality of URL $u$.
Intuitively, $A$ is the probability of one of the following, given $u$ is a viral URL ($\gamma_u$): (a)  followee $j (\alpha_{ji})$ has influenced user $i$ and tweeted $u$ with probability $p_j^u$, or (b) user $i$ tweets it under influence of an unobserved entity (or tweets spontaneously).
The time-dependent component $T$ is defined using a log-normal distribution, given complementary error function $erfc$, as
$$ T(\mu_i,\sigma_i^2,t_i^u) = \frac{1}{2}erfc(-\frac{\ln t_i^u - \mu_i}{\sigma_i\sqrt{2}}) $$
The linear threshold model (LT) they propose generalizes over ALO.
The cumulative influence from all the followees need to exceed a per-node threshold they introduce, for the user to tweet.
The $A$ component is therefore replaced by
$$ A(\alpha_{ji},\beta_i,\gamma_u) = s(\gamma_u(\beta_i + \sum\limits_{j:i \rightarrow j}\gamma_u\alpha_{ji}p_j^u)) $$
The sigmoid $s(x) = \frac{1}{1+e^{-a(b-x)}}$ serves as a continuous thresholding function.
They optimize parameters by training over using an iterative gradient ascent method, and infer the accuracy of prediction of the information (URL) cascades using {\it F-score} - the harmonic mean of {\it precision} and {\it recall}.

\cite{bakshy2012role} quantify the causal effect of social networking mediums in disseminating information, by identifying who influences whom, as well as by exploring whether individuals would propagate the same information if the social signals were absent.
They come up with two interesting findings, performing field-experiments on information sharing behavior over $253$ million subjects on Facebook, that visited the site at least once between August $14^{th}$ to October $4^{th}$, $2010$.
(a) They find that the ones exposed to given information on social media, are significantly likely to participate in propagate the information online, and do so sooner than those who are not exposed.
(b) They further show that, while the stronger ties are more influential at an individual level, the abundance of weak ties \cite{granovetter1973strength} are more responsible for novel information propagation, indicating that a dominant role is played by the weak ties in online information dissemination.

Their experiment focuses on finding how much exposure of a URL to a user is needed on their Facebook ({\it feed}) (a dashboard on the Facebook user pane, where the user is presented with information content, and a platform-level capability to share content with others), for the user to share the URL, beyond the expected correlations among Facebook friends.
Before displaying, they randomly assign subject-URL pairs to {\it feed} versus {\it no-feed} conditions, such that the number of {\it no-feed} is twice the number of {\it feed}.
Stories that are assigned the {\it no-feed} condition, but have a URL, are never displayed {\it feed}.
And the ones assigned the {\it feed} condition are displayed on the user feed and are never removed.
They measure how exposure increases sharing behavior.
They find that sharing has a likelihood of $0.191\%$ in condition of {\it feed} and $0.025\%$ in {\it no-feed}.
They note that the the likelihood of sharing is $7.37$ more for those in the {\it feed} condition.
They observe that links tend to be shared immediately upon exposure by those in the {\it feed} condition; however, those in {\it no-feed} condition share links over a marginally longer time period.
They observe that link-sharing probability goes up as more of one's contacts share a given link, under {\it feed} conditions.
On the other hand, in {\it no-feed}, a link shared by multiple friends is likely to be shared by a user, even if the user has not observed the sharing behavior of friends.
This indicates a mixture of internal influence and external correlation in information (link) sharing behavior.

The authors explore the impact of strength of ties in the diffusion of the information (URL sharing).
Studying individuals who have only one friend that has shared a link previously, they observe that both for {\it feed} and {\it no-feed} conditions, link sharing is more likely by an individual, when her friend who shared happens to be a strong tie.
This effect is seen to be more prominent in {\it no-feed}, indicating that strength of ties is a can better predict activities with external correlation than predicting influence on feed.
They observe that, ``individuals are more likely to share content when influenced by their stronger ties on their feed, and share content under such influence that they would not otherwise share''.
They further observe that the strength of weak ties \cite{granovetter1973strength} plays a significant role in consuming and transmitting information that is not likely to be transmitted and get exposed to much of the network otherwise, which increases the information propagation diversity.

\cite{yang2010modeling} propose an approach to model the ``global influence of a node on the rate of information diffusion through the underlying social network''.
To this, they propose {\it Linear Influence Model} (LIM), in which a newly infected (informed) node is modeled as a ``function of other nodes infected in the past''.
For each node, they ``estimate an influence function, to model the number of subsequent infections as a function of the other nodes infected in the past''.
They formulate their model in a non-parametric manner, transforming their setting into a simple least squares problem.
This can scale to solving large datasets.
They validate their model on 500 million tweets and 170 million news articles and blog posts.
They show that node influences are modeled accurately by LIM, and the temporal dynamics of diffusion of information are also predicted reliably.
They observe that the influence patterns of participants significantly differ with node types and information topics.

In LIM, as information diffuses, a node $u$ is treated as infected from the point of time $t_u$ that it adopts (first mentions) the information.
This enables LIM to be independent of the underlying network.
Volume $V(t)$ is defined in their setting to be the ``number of nodes mentioning the information at time $t$''.
They ``model the volume over time as a function of which other nodes have mentioned the information beforehand''.
They assign a ``non-negative influence function'' $I_u(l)$ to each node, that denotes the number of follow-up mentions, $l$ time units beyond the adoption of the information by node $u$.
The volume $V(t)$ then becomes ``the sum of properly aligned influence functions of nodes $u$, at time $t_u (t_u < t)$'':
$$ V(t+1) = \sum\limits_{u \in A(t)}I_u(t - t_u) $$
Here A(t) is the set of nodes that are ``already active (infected, influenced)''.

They propose two approaches for modeling $I_u(l)$.
In a parametric approach, they propose that ``$I_u(l)$ would follow a specific parametric form'', such as a exponential $I_u(l) = c_ue^{-\lambda_ul}$ or power law $I_u(l) = c_ul^{-\alpha_u}$, and parameters will depend upon node $u$.
They observe that the drawback of the parametric approach is that, it makes the over-simplified assumption that all the nodes would follow the same parametric form.
In a non-parametric approach, they do not assume any shape of the influence functions; the appropriate shapes are found by the the model estimation procedure.
They consider time as a discrete vector of length $L$ (a total of $L$ time slots), where the ``$l^{th}$ value represents the value of $I_u(l)$''.
To estimate the LIM model parameters, they start by marking $M_{u,k}(t) = 1$ if $k$, the contagion, reached node $u$ at time $t$, and $M_{u,k}(t) = 0$ otherwise.
Since ``volume $V_t(k)$ of contagion $k$ at time $t$ is defined as the number of nodes infected by $k$ at time $t$'', they have
$$ V_k(t+1) = \sum\limits_{u=1}^N\sum\limits_{l=0}^{L-1}M_{u,k}(t-l)I_u(l+1) $$

They subsequently rectify their model to account for the information recency (novelty) phenomenon, that nodes tend to ignore old and obsolete information and adopt recent and novel information.
To model how much more or less influential a node is when it mentions the information, they use a multiplicative factor $\alpha(t)$.
This is the $\alpha$-LIM model, represented as:
$$ V(t+1) = \alpha(t)\sum\limits_{u \in A(t)}I_u(t - t_u) $$
Here ``$\alpha(t)$ is the same for all contagions'', and is is expected to ``start low, quickly peak and slowly decay''.
They note that the ``resulting matrix equation is convex in in $I_u(l)$ when $\alpha(t)$ is fixed and in $\alpha(t)$ when $I_u(l)$ is fixed''.
Hence for estimating the ``$I_u(l)$ and $T$ values of vector $\alpha(t)$'', they apply coordinate descent, iterating between ``fixing $\alpha(t)$ and solving for $I_u(l)$, and then fixing $I_u(l)$ and solving for $\alpha(t)$''.
They also account for imitation, where everyone talks about a popular information, introducing the notion of {\it latent volume}: the volume which is caused by factors other than influence.
They add $b(t)$, a factor to model the latent volume, and thereby create the B-LIM model, which is linear in $I_u(l)$ and $b(t)$, as
$$ V(t+1) = b(t) + \sum\limits_{u \in A(t)}I_u(t - t_u) $$

\cite{yang2010predicting} explore three core properties of social network information diffusion, namely speed, scale and range.
They collect Twitter data from July $8^{th} 2009$ to August $8^{th} 2009$, for $3,243,437$ unique users and $22,241,221$ posts.
They explore the ongoing social interactions of users on Twitter, as denoted by the {\it @username} mentions (replies) and retweets, which represents active user interaction.
To measure how topics propagate through network structures in Twitter, they construct a diffusion network based on mentions.
That is, they create an edge from $A$ to $B$, if $B$ mentions A in her tweet that contains topic $C$ that $A$ had talked about earlier.
Thus, they approximate the path of person $A$ diffusing information about topic $C$.

They develop models for {\it speed}, {\it scale} and {\it range}.
For {\it speed} analysis, they attempt to understand whether and when followers would be influenced and thereby reply, retweet or otherwise mention the original tweet.
They investigate the impact of user and tweet features on the speed of diffusion, using the regression model of \cite{cox1984analysis}.
They observe that ``some properties of tweets predict greater information propagation, but user properties, and specifically the rate that a user is historically mentioned, are equal or stronger predictors''.
For {\it scale} analysis, they attempt to understand how many people in the network mentioned the same topics as the neighbors of the topic originator.
They find the number of mentions of a user to be the strongest predictor for information propagation speed (how quickly a tweet produces an offspring tweet) and scale (the number of offspring tweets a given tweet produces).
For {\it range} analysis, they trace topics through the propagation chains, and count the number of hops.
They observe that the range of information propagation (the number of social hops that information reaches on a diffusion network) is tied to the number of user mentions and when the tweets come in the observation sequence.
The tweets that come later often are seen to be more influential: those travel further over the network.

\cite{goyal2010learning} build an influence model using the Flickr social network graph and user action logs.
They propose a technique to predict the time within which a given user would be expected to conduct an action.
Other studies, such as \cite{bakshy2009social}, \cite{kimura2007extracting}, \cite{narang2013discovery} and \cite{romero2011differences}, provide significant insights into flow of information and influence, along social edges, over Twitter user interactions.
Further, other research works have attempted to model influence of content generated by users, on content generated by other users.
\cite{leskovec2007patterns}, for instance, explores bloggers' networks for modeling influence propagation.
\cite{stieglitz2013emotions} explore the correlation of sentiments that Twitter users express and their information sharing behavior, experimenting on political communication data.
From 2011 Seoul (Korea) mayoral elections data of a particular candidate who had used Twitter extensively, \cite{park2013impact} show that, rather than sharing and circulating several ideas, the communication had taken place in form of aggregation and propagation.
The communication pattern structures were fragmented rather than transitive, signifying that during the election period, the communication in general had occurred from or converged to a single node, and mostly did not circulate through multiple nodes.

\section{Topic Lifecycle on Social Media}
\label{sec:topiclifecycle}

Trend discovery from digital media text has been a research problem of significant scientific interest for long, and is still of active interest \cite{benhardus2013streaming} \cite{goorha2010discovery} \cite{kontostathis2004survey} \cite{roy2002methodologies}.
Trend and topic propagation is one of the key factors that are associated with information diffusion on online social networks.
Identifying topics and trends successfully will help in solving different practical problems.
Natural disaster analysis and recovery is one such area, explored by \cite{kireyev2009applications} \cite{mendoza2010twitter} \cite{nagar2012characterization}.
\cite{vieweg2010microblogging} empirically explore how Twitter can contribute to situational awareness, over two natural hazard events, namely Oklahoma Grassfires of April 2009 and Red River Floods of March and April 2009.
Early identification of online discussion topics of customers, can help organizations better understand and grow their products and services, as well as control damage early \cite{chen2013emerging}.
Of late, one of the key areas within this research area has been the detection of topics and trends in microblogs such as Twitter, which are often associated with one topic or a few related topics.
A number of research studies have been conducted, predominantly since $2010$, that attempt to identify trends and topics and watch them evolve and spread in social networks.

\begin{table}[thb]
\tbl{\label{tab:topiclifecycle}Literature for topic lifecycle}
{
\begin{tabular}{|p{3.7cm}|p{5.5cm}|p{3.3cm}|}
\hline
\textbf{Reference} & \textbf{Key Features / Method Overview} & \textbf{Research Outcome} \\
\hline
\cite{mathioudakis2010twittermonitor} & Detects trends on Twitter by first identifying bursty keywords occurring at sudden high rates, and then grouping these keywords into trends. Associates news sources and geographical origins of identified trends. & Produces a chart illustrating the evolution of popularity of the trend during its lifecycle. \\
\hline
\cite{lau2012line} & Proposes an online dynamic variant of LDA, that processes inputs and updates the model periodically, produces topics comparable for different periods that enables measuring the shift of topics and does not grow in size with time. & Experiments with injecting novel events on-the-fly, and shows that the model is capable of detecting topics under such settings. \\
\hline
\cite{naaman2011hip} & Creates taxonomy of geographical area-specific trends, based upon Twitter messages collected from the given areas. Identifies significant dimensions to enable trend categorization, and distinguishing features of trends. & Empirically establishes the existence of significant differences in computed features for different trend categories. \\
\hline
\cite{ifrim2014event} & Filters tweets based on the length and structure of the messages, removing noisy tweets and vocabulary. Combines with hierarchical tweet clustering, dynamic dendogram cutting and ranking of the clusters. Computes pairwise distance of tweets by normalizing the tweet term matrix and applying cosine similarity. Feeds the output into clustering. Selects the first tweet in each of the first 20 clusters as topic headlines. Re-clusters the headlines to avoid topic fragmentation. & Shows that length and structure based aggressive filtering of tweets, combined with clustering the tweets hierarchically and ranking the resulting clusters, works well for detecting and labeling events. \\
\hline
\cite{cataldi2009cosena} & Proposes a real-time emergent topic detection technique expressed by communities. Analyzes the authority of the content source using PageRank, and models term life cycles using an aging technique. & Experiments with Twitter data of $2$ days, and identifies the $5$ top emergent terms at a given time slot for demonstrating an example of their model output. \\
\hline
\cite{cunha2011analyzing} & Studies propagation and dynamic evolution of hashtags. Motivated by the concept of linguistic innovation that models language transformation, it defines hashtag innovation as a transformation of the hashtag. & Observes that individuals seeking to assign a term not yet used for this purpose for categorizing their message, tend to create new hashtags. Observes the rich-getting-richer phenomena: a few hashtags tend to attract most of the attention. \\
\hline
\cite{narang2013discovery} & Models information flow over event clusters on social media. Identifies social discussion threads by identifying social and content-based connection across event clusters, and applying temporal filters on these clusters. & Shows that topical discussions grow and evolve along social connections over time, rather than at random. \\
\hline
\cite{althoff2013analysis} & Uses historical time series data from multiple semantically similar topics to forecast the lifecycle of trending topics as they emerge. Uses nearest neighbor sequence matching, considering historical events that occurred with a similar time span. & Studies Twitter, Google, and Wikipedia, three primary online social media streams, over thousands of topics and a year, to observe the emerging trends for empirically validating their process. \\
\hline
\end{tabular}
}
\end{table}

\cite{mathioudakis2010twittermonitor} present one of the early research works in detecting Twitter trends in real-time, and analyzing the lifecycle of the trends.
They define {\it bursty keywords} as ``keywords that suddenly appear in tweets at unusually high rates''.
Subsequently, they define a trend as a ``set of bursty keywords frequently occurring together in tweets''.
Their system, TwitterMonitor, follows a two-step Twitter trend detection mechanism.
It also has a third step for analyzing the detected trends.
In the first step, they identify keywords suddenly appearing in tweets at unusually high rates, namely the bursty keywords.
In order to identify bursty keywords effectively, they propose an algorithm named {\it QueueBurst}, based on queuing theory.
The {\it QueueBurst} algorithm reads streaming data in one pass, and detects the bursty keywords in real time.
It protects against spam and spurious bursts, where, by coincidence, a keyword appears in several tweets within a short time period.

Subsequently, in the second step, they group the bursty keywords into trends, based upon co-occurrences of the keywords.
They compute a set of bursty keywords $K_i$ at every time instant $t$, that can possibly be a part of a trend (or even the same trend).
They periodically group keywords $k \in K_t$ into disjoint subsets $K_t^i$ of $K_t$, so that all keywords in the same subset are grouped under the same discussion topic.
Given subsets ${K_t^i}$, a single subset $k_i$ can identify a trend.
Thus, they identify trends as a group of bursty keywords that frequently occur together.
Identifying more keywords related with a given trend using content extraction algorithms, identifying frequently cited news sources and adding such sources to the trend description, and exploiting geographical locality attributes of the origin of tweets contributing to the identified trends (such as ThanksGiving in Canada will make it likely that a large proportion of the tweets originate from Canada), they produce a chart illustrating the evolution of popularity of the trend during its lifecycle.

\cite{lau2012line} propose a methodology for online topic modeling, for tracking emerging events for Twitter, that considers a constant evolution of topics over time, and is amenable to dynamic changes in vocabulary.
To this, they propose an online variant of the traditional LDA \cite{blei2003latent} method, which is enhanced by $(P(z|w))$, the ``posterior distribution over assignments of words to topics'', by \cite{griffiths2004finding}.
The online version of LDA they propose, processes the inputs and periodically updates their model.
It produces topics comparable across different periods, that enables measuring topic shifts.
Further, the size of topics does not grow with time.
They summarize the traditional LDA along with the incorporation of Griffiths and Steyvers \cite{griffiths2004finding} methodology as
$$ P(z = t | z, w, \alpha, \beta) \propto \frac{n(d,t)+\alpha}{n(d,.)+T\alpha} \frac{n(t,w)+\beta}{n(t,.)+W\beta} $$
Here $n(d,t)$ and $n(t,w)$ respectively denote the assignment counts of topic $t$ in document $d$ and of word $w$ to topic $t$, excluding the current assignment $z$.

To transform method into an online (streamed) one, they propose a model that can process the input and update itself periodically.
They use time-slices $k_t$, and a ``sliding window $L$ that retains documents for a given number of previous time slices''.
As time slice $k_{t+1}$ arrives, they ``resample topic assignments $z$ for all documents in window $L$'' to update the model, using the $\theta$ and $\phi$ values from the earlier model in time slice $k_t$ for serving as ``Dirichlet priors $\alpha'$ and $\beta'$ in the evolved model in time slice $k_{t+1}$''.
They introduce $c$ $(0 \leq c \leq 1)$, a contribution factor, to ``enable their model to have a set of constantly evolving topics'', where $c = 0$ indicates that the model is run without any parameter learned previously.
The time window ensures that their topic model remains sensitive to topic changes with time.
To accommodate dynamic vocabulary, they remove words falling below a frequency threshold and add new words satisfying the threshold, along time slices.

For previously seen documents and words, the ``Dirichlet priors $\alpha'$ and $\beta'$ in the new model in time slice $k_{t+1}$'' are given by: $\alpha'_{dt} = \frac{n(d,t)}{N_{old}} \times D_{old} \times T \times \alpha_0$, and $\beta'_{tw} = \beta_0 \times (1 - c) + \frac{n(t,w)}{_{old}} \times T \times W_{new} \times \beta_0 \times c$.
For new documents and words, it is calculated as $\alpha'_{dt} = \alpha_0$ and $\beta'_{tw} = \beta_0$.
Here $\alpha'_{dt}$ and $\beta'_{tw}$ are the priors for topic $t$ in document $d$ and word $w$ in topic $t$ respectively, $n(d,t)$ and $n(t,w)$ are the number of assignments in the earlier model of time slice $k_t$, and ``$D_{old}$, $N_{old}$ and $W_{new}$ are respectively the number of documents previously processed, number of tokens in those documents and vocabulary size, in time window $L$''.
They normalize to enable maintaining a ``constant sum of priors across different processing batches'', i.e., $\sum\alpha' = \sum\alpha = D \times T \times \alpha_0$ and $\sum\beta' = \sum\beta = T \times W \times \beta_0$.
For tracking events that are emerging, they measure the shifts (degree of change) in the topic model (evolution of topic) ``between the word distribution of each topic $t$ before and after an update'', using Jensen-Shannon (JS) Divergence.
If the shift exceeds a threshold, they classify a topic as novel.

They demonstrate their model using synthetic datasets on Twitter, by mixing real-life Twitter data stream (not annotated) and TREC Topic Detection and Tracking (TDT) corpus (annotated) data.
For experiments, they collect data using Twitter's streaming API from September $2011$ to January $2012$, having $12$ million tweets spanning over $1.39$ users.
They also apply their model to ``a series of Twitter feeds, to detect topics popular in specific locations''.
For experiments, the length of a time slice and window size and respectively set to 1 day and 2 days.
They find the detected popular topics to closely follow local and global news events.
They observe that, topics expressed as multinomial distributions over terms, are more descriptive compared to strings or single hashtags.
Thus, they show that their model is capable of detecting emerging topics under such settings.

\cite{petrovic2010streaming} create a locality-sensitive hashing technique, to detect new events from a stream of posts in Twitter.
Their approach is empirically shown to be an order of magnitude faster compared to the state-of-the-art, while retaining performance.
\cite{kasiviswanathan2011emerging} use dictionary learning to detect emerging topics on Twitter.
They use a two-stage approach to detect and cluster new content generated by users.
They apply their system on streaming data, showing the effectiveness of their approach.
\cite{osborne2012bieber} use the approach of \cite{petrovic2010streaming}, but filter using Wikipedia, reducing the number of spurious topics that often get detected by the topic detection systems.
They empirically show that events within Wikipedia tend to lag behind Twitter.

\cite{naaman2011hip} characterize emerging trends on Twitter.
They develop a taxonomy of geographical area-specific trends, based upon Twitter messages collected from the given geographic areas.
They denote the Twitter-given trends as $T_{tw}$.
They collect Twitter's local trending terms.
They identify the highest trending terms using a message and term frequency ($tf$) pair, such that the message set contains at least $100$ messages.
They identify bursts via terms that appear more frequently than expected in a given message set, within a given time period.
They score a term by subtracting the expected number of occurrences of all terms, from the occurrence count of the term.
They retain each term that would score in the top $30$ for a given day in a given week, for a sufficiently large number of hours.
They assemble the scores to assign a score to each bursty trend comprising of a set of such terms.
They add these terms to $T_{tw}$, and pick the top $1,500$ trends to form $T_{tf}$.

The authors run qualitative and quantitative analyses for a selective (random) subset of $T_{tw}$ and $T_{tf}$, as they observe that computing on the whole would be prohibitively expensive.
They select trends that: (a) reflect the trend diversity present in the source sets, and (b) are human-interpretable, inspecting the Twitter messages associated.
They take a set union of the trends selected, denoted as $T$, and split into two subsets, $T_{Qual}$ and $T_{Quant}$, to perform qualitative and quantitative analysis respectively.
They associate tweet messages with trends by aligning the messages with trend peak times and the surrounding $72$ hours before and after.
They observe $M_t = 1350$ in $T_{quant}$, that is, $1,350$ tweet messages on an average are associated with each trend $t$.

They broadly classify trends into two: {\it exogenous} trends that capture activities, interests and event originating outside Twitter, and {\it endogenous} trends that capture Twitter-only events that do not exist outside Twitter.
Exogenous trends comprise of global news events, broadcast media events, national holidays, memorable days and local participation-based (physical) events, while endogenous trends comprise of retweets, memes and activities of fan communities.
To characterize the two types of trends, they derive different types of features.
This includes $7$ content features based upon the content of messages in $M_t$, $3$ interaction features based upon the @-username interactions amongst users, $4$ time-based features that vary across trends and capture the temporal patterns of information spread, $3$ participation features based upon authorship of messages associated with given trends, and $7$ social network features that are built upon the followers of each message, for messages belonging to $M_t$.

They empirically establish the existence of significant differences in the set of features for different categories of trends.
They show that exogenous trends have higher URL proportions, smaller hashtag proportions, fewer retweets, fewer social connections between authors and different (temporal) head periods compared to endogenous trends.
They show that breaking news has more retweets (forwards), lesser replies (conversations) and more rapid temporal growth compared to other exogenous trends, as well as different social network features.
They notice local events to have denser social networks, higher connectivity, more social reciprocity, and more replies, compared to other exogenous trends.
They further noticed memes to have higher connectivity and more reciprocity compared to retweet trends, for endogenous events.

\cite{ifrim2014event} detect events on Twitter data of US presidential elections, collecting data from November $6^{th}$ $2012$ $23$:$30$ hours to November $7^{th}$ $2012$ $6$:$30$ hours leading to $1,084,200$ tweets, and February $25^{th}$ $2014$ $17$:$30$ hours to February $26^{th}$ $2014$ $18$:$15$ hours leading to $1,088,593$ tweets.
They base their approach on two primary steps.

One, they conduct aggressive filtering of tweets and terms, in order to remove tweets containing noise and to respect vocabulary.
They normalize tweet text and remove user mentions, URLs, digits, hashtags and punctuations.
They tokenize by whitespaces, remove stopwords, and append hashtags, user mentions and de-noised text tokens.
From the tweets thus obtained, they remove the tweets with (a) more than $2$ user mentions, or (b) more than $2$ hashtags, or (c) less than $4$ tokens.
The intuition is to eliminate tweets with too many user mentions or hashtags, but too little clean information content (text).
Effectively, this acts as noise elimination.
For vocabulary filtering, they remove user mentions, and retain bi-grams and tri-grams that are present in at least a threshold ($10$) number of tweets.
They subsequently retain tweets with at least $5$ words that are in vocabulary, in order to retain tweets that can be meaningfully clustered and eliminate tweets with little vocabulary.

Two, they combine this with hierarchical tweet clustering, dynamic dendogram cutting and ranking of the clusters.
They compute pairwise distance of tweets by normalizing the tweet term matrix and applying cosine similarity.
They perform topic-based clustering of tweets on the output using the distance thus obtained.
They cut the resulting dendogram empirically fixing at $0.5$, avoiding too tight or too loose clusters and topic fragmentation.
They rank the resulting clusters.
They observe that ranking the clusters by size, and labeling these clusters as trending topics, does not yield good results, as the topics are casual and repetitive, and by inspection appear unlikely to make news headlines.
As an alternative approach, they use the $df - idf_t$ formula of \cite{aiello2013sensing}, that approximates the current window term frequency by the average term frequency of the past $t$ time windows, as
$$ df - idf_t = (df_i+1)/(log((\sum\limits_{j=i}^t df_{i-j}/t) + 1) + 1) $$

For experiments, they set the history size $t = 4$.
They assign a high weight to the $idf_t$ term to recognized named entities, as they observe such assignments tend to retrieve more news-like topics.
They select the first tweet of each of the first $20$ ranked clusters as the headline of the topics detected.
They re-cluster the headlines to avoid topic fragmentation.
They finally present the raw tweet content of the headline (without URLs) with the earliest publication time, as the final topic headline.

\cite{cataldi2010emerging} propose a real-time emergent topic detection technique expressed by communities.
They define a term as a topic.
They define a topic as emerging if it had not occurred rarely in the past but frequently in a specified time interval.
They extract the tweet content in form of term vectors with relative frequencies.
For this, they associate a tweet vector $\overrightarrow{tw_j}$ to each tweet $tw_j$ to express all the knowledge expressed by a tweet, where each of the vector components represents a weighted term extracted from $\overrightarrow{tw_j}$.
They retain all keywords, and attempt to highlight keywords are potentially of high relevance for a topic, but appear less frequently.
Tweet vector $\overrightarrow{tw_j}$ is defined as $\overrightarrow{tw_j} = \{w_{j,1}, w_{j,2},...,w_{j,v},\}$, where $K^t$ is the corpus vocabulary in time interval $I^t$, the vocabulary size is $v = |K^t|$ and the $x^{th}$ term of vocabulary of the $j^{th}$ post has a weight $w_{j,x}$.

Based on the social relationships of active users (content authors), they define a directed graph and compute their authority using PageRank \cite{page1998pagerank}.
For each topic (term), they model the topic lifecycle using an aging technique, leveraging the authority of users, thereby studying its usage in a specific interval of time.
Each tweet provides nutrition to the contained words, depending upon the authority of the user who made the tweet.
Using keyword $k \in K^t$ and the tweet set $TW_k^t \in TW_t$ having term $k$ at the time interval $I^t$, the amount of nutrition is defined as
$$ nutr_k^t = \sum\limits_{tw_j \in TW_k^t} w_{k,j} * auth(user(tw_j)) $$
Here $w_{k,j}$ denotes the weight of the term $k$ in tweet vector $\overrightarrow{tw_j}$, the function $user(tw_j)$ gives the author $u$ for tweet $tw_j$, and the authority score of user $u$ is $auth(u)$.
Thus, they evaluate term usage frequency to quantify term usage behavior, and analyze author influence to qualify term relevance.

They formulate an age-dependent energy of a keyword using the nutrition difference across pairs of time intervals.
They define a term as {\it hot} if the term is used extensively within a given time interval, and {\it emergent} if it is {\it hot} in the current interval of time but never {\it hot} earlier.
Clearly, a keyword that has been {\it hot} over more than one time interval, then it will not be identified as {\it emergent} after the first temporal interval.
They limit the number of previous time slots to consider using a threshold.
They propose two techniques for selecting an emerging term set within a given time interval - a supervised technique and an unsupervised one.
They use a notion of {\it critical drop} \cite{cataldi2009cosena} to identify emergent topics, and proceed to label topics using a minimal set of keywords.
Critical drop is obtained as: $drop^t = (\delta . \sum\limits_{k \in K^t}(energy_k^t))/|K^t|$, where $\delta > 1$.
In a supervised setting that lets the user choose a permissible threshold for drop, they define $EK^t$, the set of emerging keywords, as: $\forall k \in K^t, k \in EK^t \Longleftrightarrow energy_k^t > drop^t$.
In an unsupervised model, they automatically set the value of this drop dynamically, by computing the {\it average drop} over successive entries for the keywords ranking higher than the maximum drop point detected, and marking the first higher-than-average drop as the {\it critical drop}.

They define topic as a ``minimal set of a terms, related semantically to an emerging keyword''.
Emerging terms are mapped to emerging topics, by studying the semantic relationships amongst the keywords in $K^t$ extracted within interval $I^t$, using co-occurrence information.
They associate a correlation vector $\overrightarrow{cv_k^t}$, defining the relationships of $k$ with all the other keywords in the interval $I^t$, in form of a weighted term set.
They create topic graph $TG^t$ using the correlation vectors, as a directed and weighted graph, where the nodes are labeled with the keywords.
Using a weight-based adaptive cut-off, they retain only the edges representing the strongest relationships, and discard the rest.

They detect emerging topics using the topological structure of $TG^t$.
For this, they discover the strongly connected components that are rooted on the emerging keyword set $EK^t$ in $TG^t$.
They define subgraph $ET_z^t(K_z,E_z, \rho)$ as the emerging topic related to each emerging keyword $z \in EK^t$.
This subgraph comprises a set of keywords, that are related to $z$ semantically, in time interval $I^t$.
$\rho_{k,z}$ represents ``the relative weight of the keyword $k$ in the corresponding vector $\overrightarrow{cv_k^t}$'' - the ``role of keyword $z$ in context of keyword $k$''.
Here, the set of keywords $K_z^t$ that belong to $ET_z^t$, the emerging topic, is obtained by ``considering as starting point in $TG^t$ for the emerging keyword $z$, but also contains a set of common terms semantically related to $z$ that are not necessarily included in $EK^t$''.
Thus they have some keywords indirectly correlated with the emerging keywords.
They rank the topics, in order to identify which topic is {\it more} emergent in the interval, as $rank_{ET_z^t} = \sum\limits_{k \in K_z^t}(energy_k^t)/|K_z^t|$.
Finally, they perform unsupervised keyword ranking, to choose the most representative keywords for each cluster.
They experiment with Twitter data of $2$ days, and identify the $5$ top emergent terms at a given time slot for demonstrating an example of their model output.

\cite{cunha2011analyzing} study dynamic evolution of Twitter hashtags.
Specifically, they investigate the creation, use and dissemination of hashtags by the members of information networks of Twitter hashtags.
They study hashtag propagation in social groups where members are known to influence each other linguistically.
They take a live and rapidly evolving content stream, and analyze the evolution of terms (hashtags).
They collect Twitter data of $55$ million users, leading to $2$ billion followership edges, out of which they find $1.7$ billion to be usable.
They compare ``features of the variation of hashtags to linguistic variation''.
They collect data from {\it interchangeable hashtags} that refer to the same event or topic, and would have been considered to be the same in a more controlled setting than Twitter.
For instance, \#michaeljackson, \#mj and \#jackson are hashtags referring to the same topic (subject).
They select topics, and form bases by filtering tweets such that a chosen tweet will have at least one hashtag, and at least one term that is well-known to be related to the topic (such as, {\it jackson} when referring to Michael Jackson).

Motivated by the concept of linguistic innovation \cite{breivik1989language} that models transformation of any language attribute such as phonetics, phonology, syntax, semantics, {\it etc.}, the authors define hashtag innovation as a transformation of the hashtag.
They observe that individuals seeking to assign a term not yet used for this purpose for categorizing their message, tend to create new hashtags; such as, to tag (name) an action or object that they are unfamiliar with in the physical (offline) world.
They observe the presence of the {\it rich-get-richer} phenomenon \cite{easley2010networks}: a few hashtags tend to attract most of the attention, with only around $10\%$ of the hashtags getting used more than $10$ times, and as many as $60\%$ of the hashtags getting used only once.
They observe that hashtags that gain the maximum popularity tend to be direct, short in length and simple, while many of the less popular hashtags are formed by long character strings.
They also clearly observe that the difference in lengths of the top few popular tags are irrelevant.
However, they compare between the more popular and less popular hashtags and conclude that the number of characters in a given hashtag, a linguistic (and internal) feature, determines the success/failure of the hashtag on Twitter.

\cite{narang2013discovery} model information flow over topics on social media, using empirical evidence found from natural disaster and political event datasets of Twitter.
They introduce the notion of social discussion threads by creating event clusters on Twitter data, connecting across these clusters based upon contemporary external news sources about the events under consideration, and examining the social and temporal relationships across cluster pairs.
They identify conversations by exploring social, semantic and temporal relationships of these clusters.
Their model also looks at temporal evolution of the topics as they evolve in the social network, over discussions.

They represent an event as $E^i = \{(K_1^i, K_2^i, ..., K_n^i),[T_s^i,T_e^i]\}$, where $K^i$ is the keyword set extracted from the tweets belonging to event $E^i$, and $T^i$ is the event time period.
$K$ contains the proper nouns (extracted using PoS tagging) and $idf$ vector from the tweets.
Thus, each event becomes a cluster of tweet messages.
They define extended semantic relationships across event cluster pairs, connecting the pairs with information obtained from contemporary external document corpus such as Google News.
They generate $|K^i| \times |K^j|$ keyword pairs that need to be evaluated for extended semantic relationship, pruning semantically related pairs such as synonyms, antonyms, hypernyms and hyponyms in order to avoid skewed results.
They use the Wordnet lexical database to compute similarity of keyword pairs, and retain keyword pairs with sufficient similarities.
They find contemporary external documents in which both the keywords occur.
They compute a document pair coupling score, such that, ``if $C(K_l^i, D_t)$ is the {\it tf-idf} score of word $K^i$ in document $D^t$, the pairwise coupling score is given as $minimum(C(K_l^i, D_t), C(K_m^j, D_t))$''.
They calculate the coupling score of a pair of keywords as the average coupling score across all documents.
Extending this to all keyword pairs for a given event cluster pair $E^i$ and $E^j$, if $w_{ij}$ keyword pairs were retained and the rest were pruned, they compute the overall score of connection of the event pair as
$$ \text{Overall score} = \frac{\sum\limits_{K^i, K^j}Coupling}{(|K^i| \times |K_j|) - w_{ij}} $$

In their setting, a person $P$ belongs to an event cluster $E^i$, iff $P$ posts a message $M$, such that $M \in E^i$.
This allows a person to belong to multiple clusters simultaneously.
An edge is created between clusters $E^i$ and $E^j$, if person $P_i \in E^i$, $P_j \in E^j$, and $(P_i, P_j)$ is a social followership edge in the input Twitter graph.
If $E^i$ and $E^j$ have $P^i$ and $P^j$ memberships respectively, the average neighbor count in $E^j$ ($E^i$) of an individual in $E^i$ ($E^j$) is $a_{ij}$ ($a_{ji}$), then the edge $(E^i, E^j)$ has a strength of $P^i.a_{ij} + P^j.a_{ji}$.
They create two kinds of temporal relationships across event cluster pairs, drawing from Allen's temporal relationship list \cite{allen1983maintaining}.
They create a ``temporal edge from event $E^i$ to event $E^j$, if $E^j$ starts within a threshold time gap after $E^i$ ends'', and set this gap to $2$ days for experiments, and label as {\it follows}.
This is effectively the set union of Allen's {\it meet} and {\it disjoint} relationships.
They also create the temporal {\it overlap} relationship of Allen across cluster pairs.

They propose a two-step process for identifying social discussion threads that evolve topically.
First, they construct the {\it semantic AND temporal} graph by taking edge set intersection of event cluster pairs, considering direction, to form discussion sequences.
Next, they construct the {\it semantic AND temporal AND social} graph, by also intersecting the social edges.
This retains the socially connected discussion sequences, and discards the others, thereby identifying social discussion threads.
They extract modularity-based communities from the discussion sequences as well as the social discussion threads, and find the normalized mutual information (NMI) \cite{coombs1970mathematical} of the two.
Over multiple datasets, they show that this NMI value is significantly higher, compared to the NMI value across the communities found in the input social and semantic graphs.
They claim this as evidence of topical discussions growing and evolving along social connections over time, rather than at random, even for events of large scale where randomness of user participation and discussion is likely.
They also qualitatively show that discussion threads tend to localize in social communities.

\cite{althoff2013analysis} propose an approach to forecast the life cycle of trending topics as they emerge.
They observe popular terms from $10$ different sources, including $5$ Google channels, $3$ Twitter channels and $2$ Wikipedia channels.
Retrieving $10$-$20$ feeds per day (total $110$ topics per day), they observe over thousands of topics and a period of a year.
They unify the trends found across different sources using edit distance clustering.
They rank each trending topic (cluster) by assigning a global trend score as the sum of daily trend scores.
They define lifetime of a trend as ``the number of consecutive days with positive trend scores''.
Doing lifetime analysis of trends, they investigate the survival duration of trends, its variation across different media channels.
They find trends to last typically less than $14$ days.
They observe that Twitter trends to be the shortest, and Wikipedia trends also to be short.
They observe Google to cover a significant proportion of the major trends, and thus Google dominates the lifetime histogram of the topics that trend.
They observe that certain categories of topics go well with certain channels.
For instance, sports is the most popular on Google, while holidays, celebrities and entertainment are most popular on Twitter.

Using historical time series data from multiple semantically similar topics, they forecast which of the emerging topics will trend.
This comprises of three steps.
First, they discover semantically similar topics.
They use DBPedia \cite{auer2007dbpedia} named entities and category information.
They create a topic set that includes all discovered similar topics.
To find similar topics, they define two topic sets: one including the trending topic, and another containing various general topics (to compare with the trending topic).
Second, they do a nearest neighbor sequence matching, on time-series of topics of interest, using ``the viewing statistics of the two previous months, to all partial sequences of same length of similar topics in the set of topics''.
Third, they forecast the life cycle of trending topics.
Their forecast draws from the best matching semantically similar topic.
It uses the semantic similarity score to ``scale to adjust to the nearest neighbor time series''.

\cite{makkonen2004simple} propose incorporating simple semantics into topic detection for documents, by grouping the terms based upon similar meanings.
They associate the group with external ontology, and extract terms and entities into distinct sub-vectors to represent the document.
Similarity of a given pair of documents are computed using sub-vector similarity.
\cite{lu2012trend} predict topics that would draw attention in future.
They use {\it moving average convergence divergence} (MCAD), an indicator frequently used to study stock prices, to identify emerging topics, using a short-period and long-period trend momentum oscillator, and average of term frequency.
They predict that a term will trend positively if a trend with a negative momentum changes to positive, and will trend negatively if a trend with a positive momentum changes to negative.

\section{Topic Dynamics and Familiarity/Similarity Groups}
\label{sec:famsim}

Bringing the aspects of familiarity and similarity together, finding the impact of one on the other, and correlating the two for information modeling, have drawn research interest.
Research questions that require study of familiarity and similarity of users of online social networks have been asked, such as whether topics of interest are more similar among users with following relations that without, and whether recommending a user to make a social connection with another user based upon similarity is effective.
In Twitter, homophily \cite{mcpherson2001birds} implies that a ``user follows a friend if she is interested in one or more topics posted by the friend, and the friend follows her back because she finds that they share similar topical interest(s)''.
Researchers have investigated homophily for information diffusion and community analysis.

\begin{table}[thb]
\tbl{\label{tab:familiarsimilartopicalcommunities}Literature for social familiarity and similarity for information modeling}
{
\begin{tabular}{|p{2.5cm}|p{5cm}|p{5cm}|}
\hline
\textbf{Reference} & \textbf{Key Features / Method Overview} & \textbf{Research Outcome} \\
\hline
\cite{weng2010twitterrank} & Investigates the presence and causes of reciprocity in Twitter followership network, and impact of this reciprocity. & Shows that Twitter users with reciprocal followerships are topic-wise more similar, compared to those without. Shows that Twitter followerships are more interest-based than casual. \\
\hline
\cite{gupta2013wtf} & Proposes SALSA, a user-recommendation stochastic algorithm for a user to follow other users, based upon user-expressed interest and the set of people followed. This lets their system recommend other users to a given user. & Observes that users who are similar often follow one another, and users often follow other users that in turn follow similar other users. \\
\hline
\end{tabular}
}
\end{table}

In one of the earliest works, \cite{weng2010twitterrank} attempt to bring social familiarity and similarity together in social network and microblog settings.
They collect data for $996$ top Twitter users from Singapore in terms of number of followers, as per {\it twitterholic.com}.
They crawl the followers and friends (those being followed) of each of these users $s \in S$, and store them in the set $\bar{S}$.
They finalize their set of target users for the experiment as $S' = S \cup \bar(S)$.
Thereby, they obtain $S^* = \{s|s \in S' \text{, and } s \text{ is from Singapore }\}$.
In their data, $|S^*| = 6748$.
They represent the set of all tweets by all members of $S^*$ by $T$, where $|T| = 1,021,039$ for their dataset.
They observe that, except for a few outliers, the number of tweets made the the users, the number of followers and the number of friends (those being followed), follow the power law distribution.
They observe that the Twitter platform is rich in the reciprocity property: in spite of an edge (followership) being a one-way relationship, ``$72.4\%$ of Twitter  users follow back more than $80\%$ of their followers, and $80.5\%$ of the users have $80\%$ of users they follow, following them back''.

To determine the presence of homophily on Twitter, they ask whether topics of interest are more similar among users with following (and reciprocal following) relationships compared to those without.
To answer, they attempt to find topic interests of Twitter users, since topics are not explicitly specified on Twitter, and hashtags are not present in all messages.
They collect all tweets made by a user, and create a user-level document, and repeat this for each user.
They run LDA \cite{blei2003latent} for topic detection.
In the LDA process, they create $DT$, a $D \times T$ matrix, where $D$ and $T$ respectively denote the count of users and topics.
$DT_{ij}$ represents the ``number of times a word in user $s_j$'s tweets is assigned to topic $t_j$''.
They measure {\it topical difference} between a pair of users $s_i$ and $s_j$ as
$$ dist(i,j) = \sqrt{2 * D_{JS}(i,j)} $$
The JS Divergence $D_{JS}(i,j)$ between probability distributions $DT'_i$ and $DT'_j$ is calculated as
$$ D_{JS}(i,j) = \frac{1}{2}(D_{KL}(DT'_i.||M) + D_{KL}(DT'_j.||M)) $$
Here ``$M$ is the average of the two probability distributions, and $D_{KL}$ is the KL divergence \cite{kullback1951information} of the two''.
Using the notion of {\it topical difference}, they perform statistical hypothesis testing and find in answer to their question, that, users with following (and reciprocal following) are more similar in terms of topics of interest, than those without.

They attempt to measure topic-sensitive influence of Twitter users, by proposing a PageRank-like \cite{page1998pagerank} framework, and call it {\it topic-specific TwitterRank}.
They consider the directed graph, where edges are directed from followers to friends (persons followed).
They perform a topic-specific random walk, and construct a topic-specific relationship network among Twitter users.
For a topic $t$, the random surfing transition probability, from follower $s_i$ to friend $s_j$, is defined as
$$ P_t(i,j) = \frac{|T_j|}{\sum\limits_{a: s_i \text{ follows } s_a}|T_a|} * sim_t(i,j) $$
Here $s_j$ has published $|T|$ number of tweets, and $\sum\limits_{a: s_i \text{ follows } s_a}|T_a|$ is the total number of tweets published by all the friends of $s_i$.
The similarity between $s_i$ and $s_j$ in topic $t$ can be found as
$$ sim_t(i,j) = 1 - |DT'_{it} - DT'_{jt}| $$

This definition captures two notions.
(a) It assigns a higher transition probability to friends who publish content more frequently.
(b) The influence is also based upon topical similarity of $s_i$ and $s_j$, capturing the homophily phenomenon.
They introduce measures to account for pairs of users that follow only each other and nobody else.
For this, they use a teleportation vector $E_t$, which captures the probability of a random walk {\it jumping} to some users rather than following the graph edges all the time.
They calculate topic-specific TwitterRank $\overrightarrow{TR_t}$ of users, in topic $t$, iteratively as $\overrightarrow{TR_t} = \gamma P_t \times \overrightarrow{TR_t} + (1-\gamma)E_t$, where $P_t$ is the transition probability and $\gamma$ ($0 \leq \gamma \leq 1$) controls the teleportation probability.
The TwitterRank vectors thus constructed are topic-specific
They capture the influence of users for each topic, and aggregate to compute the overall influence of users, as
$$ \overrightarrow{TR} = \sum\limits_t r_t . \overrightarrow{TR_t} $$
Here topic $t$ is given weight $r_t$, and the corresponding $\overrightarrow{TR_t}$.
Weight assignments differ across different settings, to compute user influence under such settings.
Their study reveals that the high reciprocity in Twitter can be explained by homophily.
This empirically shows that Twitter followerships are more interest-based than casual.

\cite{gupta2013wtf} observe that on Twitter, a user tends to follow those who are followed by other similar users.
Thus, the followers of a user tend to be similar to each other.
They claim that user similarity is likely to lead to followership (familiarity).
Motivated by this, they deploy a few user recommendation algorithms (a user recommended to another user for followership) in Twitter's live production system.
One algorithm is based upon user's {\it circle of trust}, derived from an egocentric random walk similar to personalized PageRank \cite{bu2010haloop} \cite{fogaras2005towards}.
The random walk parameters include the count of steps, reset probability (optionally discarding low-probability vertices), control parameters used to sample outgoing edges for high outdegree vertices {\it etc}.
They dynamically adjust the random walk and personalization parameters for specific applications.

They deploy another algorithm based upon SALSA (Stochastic Approach for Link-Structure Analysis) \cite{lempel2001salsa}, a random walk algorithm like PageRank \cite{page1998pagerank} and HITS \cite{kleinberg1999authoritative}.
SALSA is applied on a hub-authority bipartite graph such that it traverses a pair of links at each step, one forward and one backward link.
This ensures that the random walk ends up on the same side of the bipartite graph every time.
For each user, the hub comprises of a set of users that a given user trusts, and the authority comprises of a set of uses that hubs follow.
They run SALSA for multiple iterations and assigns scores to both the sides of the bipartite graph.
On one side of the bipartite graph, they obtain a {\it interested in} kind of rank of the vertices.
On the other side, they obtain user similarity measures.
This lets their system recommend other users to a given user, using a rank of similarity of users that are thus reached in the random walk process, where the ranks are computed based upon expressed interest, and the set of people followed.
They evaluate on Twitter using offline experiments on retrospective data, as well as A/B split testing on live data, and find SALSA the most effective among the different follower recommendation algorithms for Twitter.

Among other studies that involve social familiarity and similarity, \cite{ying2010mining} model social network user similarity using trajectory mining.
\cite{rad2013similarity} analyze YouTube social network user communities and apply several measures of similarity on the communities.
Some of the similarity computation methods they apply include Jaccard \cite{jaccard1901etude} and Dice \cite{dice1945measures} similarity co-efficient, Sokal and Sneath similarity measure \cite{sneath1973numerical}, Russel and Rao similarity measure \cite{russell1940habitat}, Roger and Tanimoto similarity measure \cite{rogers1960computer} and $L^1$ and $L^2$ norms \cite{gradshteyntables}.
They observe that communities are formed from similar users on Youtube; however, they do not find the friends in YouTube communities to be largely similar.

\cite{modani2014like} attempt to find like-minded communities on a movie review platform that also has a social network friendship platform inbuilt.
They define like-mindedness as a measure to capture the compatible interest levels among community members, as cosine similarity of ratings the members assign to different movies.
They find communities with the objective being like-mindedness.
Using frequent itemset mining, they find tight small groups with multiple shared interests, that act as core building blocks of like-minded communities.
Comparing with communities discovered using only interaction information, they show these communities to have higher similarity of interests.

\cite{aral2009distinguishing} attempt to distinguish between influence-based diffusion and homophily-driven contagion in product-adoption decisions, on dynamic networks.
They investigate the diffusion of a mobile service product for $5$ months after launch, on the Yahoo instant messenger (IM) network, a social network that comprised of 27.4 million users at the time of experimentation.
They use a dynamic match framework for sample estimation that they develop to differentiate influence and homophily effects in a dynamic network setting.
Their findings indicate that ``homophily explains more than $50\%$ of perceived behavioral contagion''.
While this study is not a direct investigation of impact of familiarity on similarity or vice-versa, it is one of the early works on social networks that show the significance of similarity (homophily) on a social network, and contrast this with the impact of peer influence.
\cite{de2010birds} consider similarity and social familiarity together, investigating the impact of homophily on information diffusion, as outlined in Section~\ref{sec:topicalflowdiffusion}.

Many research works exist that address similarity and familiarity independently.
Different kinds of similarities between users have been studied on social networks and microblogs, like Facebook and Twitter.
Early studies attempted to measure tag-based similarity of users.
For instance, \cite{laniado2010making} measure user similarity based upon Twitter hashtags.
Topic-based similarity of users refines the notion of tag-based similarity of microblog users.
\cite{hong2011predicting} propose to train topic models using two different methodologies: LDA \cite{blei2003latent} and author-topic model \cite{rosen2010learning}.
They subsequently infer topic mixture $\theta$ both for corpus and messages.
They ``classify users and associated messages into topical categories'', to empirically demonstrate their system on Twitter.
They use JS divergence to measure similarity between topics.
Based upon this, they classify users into topical categories, which in turn can act as a foundation for measuring similarities of user pairs.
In a study focusing on Twitter user sentiments (opinions), \cite{tan2011user} empirically show that, under the hypothesis that connected (familiar) persons will have similar opinions, relationship information can complement what one can extract about a persons's viewpoints from their explicit utterances.
This in turn can be used to improve user-level sentiment analysis.

\section{Geo-Spatial Topical Communities and Their Evolution}
\label{sec:spatiotemporal}

Different topics on social networks receive different levels of visibility and traction at different geo-locations.
Further, the span of these topics, from inception of a topic to the topic passing through its lifecycle, vary across geographies, depending upon the nature and the locality of the events.
Spatio-temporal analysis of microblog topics and modeling topical information diffusion in spatio-temporal settings are active research areas.
Several works have attempted to analyze spatio-temporal aspects of social media and microblogs, mostly Twitter, with different angles of application.

\begin{table}[thb]
\tbl{\label{tab:geotemporalcommunityevolution}Literature for spatio-temporal evolution of topic-based social information diffusion}
{
\begin{tabular}{|p{2.5cm}|p{5cm}|p{5cm}|}
\hline
\textbf{Reference} & \textbf{Key Features / Method Overview} & \textbf{Research Outcome} \\
\hline
\cite{ardon2013spatio} & Characterizes spatio-temporal characteristics of diffusion of ideas on Twitter. Investigates network topology of followership and geo-spatial location of users on user graphs discussing a given topic. & Shows that topics become popular if the follower count of the topic initiator is high and the topic is received by users having just a few followers. Shows that popular topics cross geographical boundaries, and disjoint clusters of popular topics merge to form a giant component. \\
\hline
\cite{nagar2013topical} & Identifies and characterizes topical discussions at various geographical granularities. Assigns users and tweets to locations, and creates temporal and geographical relationships across event message clusters, thereby identifying discussions. & Observes geographical localization of temporal evolution of topical discussions on Twitter. Finds discussions to ``evolve more at city levels compared to country levels, and more at country levels compared to globally''. \\
\hline
\cite{lee2011novel} & Analyzes spatio-temporal dynamics of user activity on Twitter. Creates a two-pass process: a content and temporal analysis module to handle micro-blog message streams and categorize them into topics, and a spatial analysis module to assign locations to the messages on the world map. & Observes that the distribution of users who discussed a given event becomes global once a news media broadcasted a given news, expanding the geographical span of the locations associated with the event. Recognizes an event as a local one if it has a distribution of a high-density, and global otherwise. \\
\hline
\end{tabular}
}
\end{table}

\cite{ardon2011spatio} \cite{ardon2013spatio} conduct some of the pioneering studies to characterize the spatio-temporal characteristics of diffusion of ideas on Twitter.
On the subgraphs that form out of users discussing each given topic, they study two time-evolving properties: network topology of followership and geo-spatial location of users.
They use Twitter data collected between June and August $2009$, spanning over $10$ millions users and $196$ millions tweets.
They infer geo-locations from GPS data and user-specified data on Twitter in form of latitude-longitude pairs, using Yahoo! Placefinder service API to resolve in terms of city, state and country.
They take the hashtags as topics.
Since only $10\%$ of the tweets have a hashtag in their dataset, they also augment the set of topics by tagging tweets with entities, topics, places and other such tags, extracted using a text analytics engine (OpenCalais), and allowing a tweet to have multiple tags.

They use the term {\it event} for major or minor happenings causing surge in tweeting activity of a given topic.
In their model, they partition events into five divisions: {\it pre-event phase} when a topic gets initiated in the social network, {\it growth phase} when the topic is discussed by early adopters, {\it peak phase} when the topic is discussed by an early majority of individuals, {\it decaying phase} when the topic is discussed by a late majority of individuals and {\it post-event phase} the topic is discussed by laggards.
They experiment with three event categories they created to perform the characterization: ``{\it popular events} having $10,000+$ tweets, {\it medium-popular events} having between $500$ and $10,000$ tweets and {\it non-popular events} having between $100$ and $500$ tweets''.
For each topic, they construct a subgraph ({\it lifetime graph}) of individuals who, at any time in the window, have tweeted at least once on the topic.
They investigate a {\it cumulative evolving graph} for a topic, that captures the cumulative action of a user tweeting on the topic on at least one given day.
They also study an {\it evolving graph} for a topic, which captures the action of a user tweeting on the topic on a given day.

Analyzing the above graphs, they observe that popular topics aggressively cross regional boundaries, but unpopular topics do not.
They hypothesize that popularity and geographical spread of topics are correlated.
They count the number of regions with at least one individual mentioning a topic and plot it against the topic's popularity.
The plot indicates that popular topics typically touch a higher number of regions compared to the less popular ones.
In order to prove their hypothesis, in the {\it cumulative evolving graphs}, they compute the proportion of edges $(u \rightarrow v)$ for each topic, such that $u$ and $v$ belong to two different geographical regions.
They observe that the fraction of edges that cross boundaries of geographies throughout their evolution, is high for the {\it popular events} ranging from $0.74$ to $0.81$ in their experiments.
This fraction is observed to be low in case of {\it medium-popular events}, and very low in case of {\it non-popular events}.
In summary, this part of their analysis shows that, the more popular a topic is on Twitter, the higher will be the fraction of edges crossing geographical boundaries, across all temporal phases on the event in its lifecycle of existence.

Analyzing $4,000$ popular and less-popular topics, they show that, a large, connected subgraph tends to be formed by most users, discussing some popular topic on a given day.
However, discussions on less popular topics tend to be restricted to disconnected clusters.
They infer that ``topics become popular when disjoint clusters of users discussing them begin to merge and form one giant component that grows to cover a significant fraction of the network''.
They find the popularity of a given topic to be high, where the number of followers of the topic initiator is high.

\cite{nagar2013topical} conduct a geo-spatial analysis of topical discussions on unstructured microblogs, empirically demonstrating on Twitter.
They identify and characterize topical discussion threads on Twitter, at different geographical granularities, specifically countries and cities.
They cluster the tweets based upon topics, and draw the notions of extended (contextual) semantic and temporal relationships, from \cite{narang2013discovery}.
They create geographical relationships across pairs of clusters based upon the geo-location that the constituent tweets and users belong to.
In order to compute geographical relationships, they assign users and tweets to locations with certain probabilities, based upon the users profiles and tweet origins.
They propose two definitions of {\it belongingness} of a cluster to a geographical region: one based upon the geographical distribution of users whose messages are included in the cluster, and the other based upon the geographical distribution of origination of the tweets that constitute the cluster.

They extract geographical relationships at two granularities: cities and countries.
Given location $L^i$ and event cluster $E^i$, $L^i \in E^i$ iff at least one microblog post $M_i$ is made from a location in $L^i$, such that $M^i \in E^i$.
This allows a location to be a part of multiple clusters at the same time.
Each event cluster thus gets a vector of locations $L^i = (L_1^i, L_2^i, ..., L_m^i)$ associated to it.
For each location associated with a cluster, they compute a {\it belongingness} value of the cluster to the location.
This gives a belongingness value vector.
They quantify the geographical relationship strength for each cluster pair, by associating geographies with the each of the belongingness value vectors.

To compute {\it belongingness}, they augment the $L^i$ vector to a $\hat{L}^i$ vector, where each element $\hat{L}^i_c \in \hat{L}^i$ is assigned to a belongingness value $\hat{B}^i_c$.
Here $c$ the index of an individual location within the location vectors $L^i$ and $\hat{L}^i$, and $1 \leq c \leq m$.

They assign values to each belongingness vector $\hat{B}^i$ elements using two methods: ``the count of distinct individuals belonging to each location'', and ``the tweet count originating from each location''.
Given $|S^i_p|$ total distinct members in the event cluster $E^i$, out of which $|\hat{S}^i_p|$ distinct members belong to location $\hat{L}^i_c$, they quantify the {\it belongingness} value of $\hat{L}^i_c$ to $E^i$, by the participating user measure, as $\hat{B}^i_c = |\hat{S}^i_p|/|S^i_p|$.
When computed using the number of tweets $|\hat{S}^i_p|$ from location $\hat{L}^i_c$ in event cluster $E^i$, and the total number of tweets $|S^i_t|$, they quantify the value of {\it belongingness} as $\hat{B}^i_c = |\hat{S}^i_t|/|S^i_t|$.

They subsequently connect (relate) event clusters $E^i$ and $E^j$ with geographical relationship edges, that have edge weights given by fuzzy similarity of the sets as: $W_{ij} = \hat{B}^i_c.\hat{B}^j_d/max(\hat{B}^i_c.\hat{B}^i_c,\hat{B}^j_d.\hat{B}^j_d)$.
Here $\hat{B}^i$ and $\hat{B}^j$ represent the respective belongingness value vectors of $E^i$ and $E^j$.
Further, $1 \leq c \leq m$, $1 \leq d \leq m$, $\hat{B}^i_c.\hat{B}^j_d \neq 0$ iff $c \neq d$, ensuring that they can apply the formula of set similarity on these vectors.
They treat the dot $(.)$ as the simple multiplication operator in their experiments.
Thus, they construct a geographical relationship graph.

They create temporal relationships also across these clusters, and find geo-social discussion threads for given topics, as well as the evaluation of these topics of discussions.
They draw the notion of discussion sequences from \cite{narang2013discovery}, and define a geographical discussion sequence as the edge set intersection of the discussions sequences and the geographical relationship graph, over the same set of vertices (event clusters).
They use improvement of NMI \cite{coombs1970mathematical} between the discussion sequence graph and the geographical discussion threads graph, over the raw pairs of input semantic and geographical relationship graphs, to establish the goodness of their method.
Experimentally, they observe geographical localization of temporal evolution of topical discussions on Twitter.
Their findings indicate that the evolution of topical discussed on finer geographically granularities are stronger that coarser granularities.
They empirically find that discussions tend to evolve the most at city levels, comparatively lesser at country levels, and the least at a global level.

\cite{lee2011novel} attempt to understand how to use social networks as a spatio-temporal source of information, conducting a spatio-temporal analysis of user activity dynamics on Twitter.
They develop algorithms to mine microblogging text stream in order to obtain geo-spatial event information in real time, and thereby detect and group emerging topics.
They create a ``two-pass process over two modules: a content and temporal analysis module, and a spatial analysis module''.

In the first pass, they categorize the microblog message streams into topics.
They define an {\it event} as ``a set of highly concentrated messages focused on some issues in a given period of time, which is also described as the characteristics of temporal locality among messages''.
Given a message stream $M = \{m_1, m_2, ..., m_k, m_{k+1}, ...\}$ (temporally ordered) that arrive at time $T = \{t_1, t_2, ..., t_k, t_{k+1}, ...\}$, $\forall m_k, m_j \in M, \nexists m_k = m_j$, they remove messages having non-ASCII characters by applying a language filter, and construct a bag-of-words from the text.
They construct a dynamic feature space using a sliding window model, to handle streams of messages.
They assign weights dynamically to each word using a dynamic terms weighting method that compares historical records.
They establish relationships across messages using a neighborhood generation algorithm, and use DBScan for text stream clustering.
They thus continuously group messages into topics.
The cluster shapes keep changing over time.
They analyze the clusters to determine the hot topics from the posts.

In the second pass, the spatial analysis, they assign locations to the messages on the world map, using the spatial locality characteristics of messages.
Spatial locality of messages describes the high concentration of a set of messages in a specific geo-location.
They record the distribution of location of topics at a given point of time using a location feature vector.
They observe that the distribution of the population that discussed a given event would become global once a news media broadcasts a given news, expanding the geographical span of the location feature vector associated with the detected event.
They formulate the probability of topic $topic_t$ belonging to location $loc_j$, as
$$ p(L = loc_j|topic_i) = \frac{|occur_{i,j}|}{N_i}*\frac{1}{|loc_j \in topic_i|} $$
In other words, they derive the probability of $topic_t$ belonging to location $loc_j$ as the ratio of the message count containing $loc_j$ in $topic_i (occur_{i,j})$ to the total message count $N_t$.
Topics discussed widely across many locations are penalized with a penalty factor $1/(|loc_j \in topic_i|)$.
They determine a candidate location by the maximum for probability of $topic_i$ as: $candiLoc(topic_i) = argmax_{loc_j}\{p(L = loc_j|topic_i)\}$.
They compute whether a topic would be recognized as local or global, as
\[
	Loc_i = 
	\begin{cases}
		candiLoc(topic_i) & \text{if } p(L = candiLoc(topic_i)|topic_i) > \theta \\
		\text{\it ``globalTopic''}	& \text{otherwise}
	\end{cases}
\]
The sparsity level and the concentricity of a given topic are traded off using a cut-off point $\theta$.
The authors note that a topic remains local if the likelihood of a candidate location crosses the threshold point.
Thus, they recognize an event with high distribution density as local, and otherwise as global.
They experimentally demonstrate the effectiveness of their method, over $52,195,773$ Twitter messages collected between January $6^{th}$ $2011$ and March $11^{th}$ $2011$.

In a study that demonstrates the real-life effectiveness on pandemic disease data that authorities use for disease control, \cite{achrekar2011predicting} use Twitter data to collect related hashtag-based data pertaining to influenza-like diseases.
Using user's known position (such as from 3G phone) and profile location and periodically collected data, form a spatio-temporal influenza database.
Their experiments show a high ($0.9846$) correlation coefficient with ground-truth illness data reported to the authorities.
They use this platform to develop a regression model, that effectively improves predicting influenza cases.
\cite{singh2010situation} analyzes a combination of geo-spatial and temporal interest patterns on Twitter, for situation detection and control applications, from text, image and video data.
They demonstrate the effectiveness of their system on a Swine Flu monitoring application.
\cite{zhang2012beyond} observe the presence of meaningful temporal, geo-spatial and geo-temporal tag clusters on Flickr dataset.
To enable easy recognition of semantic relationships across tags by humans, they provide a visualization system for geographical and temporal tag distributions.
\cite{noulas2011empirical} analyze check-in behavior and inter-checkin distances of users to several geo-locations, using spatio-temporal patterns in user mobility.
They also analyze activity transactions: find a likely next activity given a current activity at a location.
\cite{lee2010measuring} detect unusual geo-social events from Twitter, using geo-tagged tweets and geographical regularities from the usual crowd behavior patterns, and finding deviations from these patterns at the time under consideration.

\section{Discussions}
\label{sec:discussions}

In the current article, we conducted an in-depth study of representative state-of-the-art models found in the social network analysis literature for topic detection and evolution, information diffusion, and properties of social connections.
The scope of the article permitted us to cover some of the social network information diffusion models, including ones that factor the impact of influence and topics, the lifecycle, geo-spatial spread and temporal evolution of topics and their role in diffusion of information, the social structures that emerge out of content and topics from the underlying social connection graph, and the mutual impact of familiarity and similarity in this context.
While plenty of research has been conducted to enrich a number of the areas, there are still many shortcomings and open problems that need to be addressed.
Some of the problems are discussed below.

\subsection{Detection and Spread of Topics}
\label{subsec:toppopdet}
Topics are identified using: (a) hashtags of microblogs like Twitter (ex: \cite{cunha2011analyzing}), (b) bursty keyword identification (ex: \cite{cataldi2010emerging} and \cite{mathioudakis2010twittermonitor}), and (c) probability distributions of latent concepts over keywords in user generated content (ex: \cite{lau2012line}).
Bursty topics are often treated as trending topics for modeling and analysis.
The shortcomings in topic detection related literature appear to be the following.

{\bf Consideration of social influence:}
Literature exploring the impact of influence on emergence of topics leaves many questions open.
A better understanding is needed on, whether users having general and topic-specific influence create long-lasting topics and high information outreach.
How do structures such as communities emerge from social connections?
What is the role of influences around topics there?
Do topics created by different influencers tend to spread together or compete with each other?
What is the social relationship of influencers in such setting?

{\bf Managing topic complexity along with scale of detection:}
Hashtags and bursty keywords, two of the popular methods to identify topics/trends, often represent simple single-word concepts.
These are often not disambiguated, leading to information loss.
For instance, \#IITDelhi and \#IITDelhiIndia are conceptually the same ``topics'' (or trends), and yet mostly treated as different topics in literature.
No work unifies such concepts automatically (\cite{cunha2011analyzing} unifies manually).
Algorithms to detect topics as probability distributions over n-gram concept sets do not scale enough to cover a large enough fraction of social network messages fast.
Identifying complex topics fast and at scale, while representing without information loss, needs research focus.

{\bf Information-rich multimedia data analysis:}
There is space to improve the state of the art of topic detection, by considering not just text but also other kinds of inputs such as images and videos, for detection topics of interest and thereby conducting analyses.
One could also consider the commonalities of the types of resources shared, such as objects that the URLs shared by the users point to, in order for topic detection.
The existing literature has not explored this.

{\bf Consideration of state of the social network:}
Topics may not necessarily emerge from external events.
Topics might get created because of the state that a given social network already is in.
This is not yet explored in the literature.
In such settings, the state of the social network can be determined by the prior set of topics, ongoing discussions, set of participants, their social relationships and other relevant attributes, and be filtered via aspects such as geographies and communities.

{\bf Defining Discussions:}
The literature mostly assumes that a microblog discussion is nothing but a topic (such as a Twitter hashtag) being mentioned by members of a social network, without attempting to define discussions and validate any such definition.
Some research works, such as \cite{narang2013discovery}, attempt to define discussions using message clustering and temporal filters.
However, attention is clearly required to better define discussions, and justify such definitions.

{\bf The closed-world assumption:}
Literature usually treats topic lifecycle and information diffusion as incidents internal to given social networks, as a closed world.
However, a preliminary study by \cite{myers2012information}, shows significant impact of external information sources, on information diffusion.
This necessitates conducting a deeper study of external impact on information diffusion, and exploring the validity of the closed world assumption that most of the literature assumes.

\subsection{Evolution of Topics}
\label{subsec:eovoftop}
Topics evolve with time, participants and geography.
Literature attempts to model the evolution and lifecycle of topics.
Some of the scope of improving the state of the art in this space is outlined below.

{\bf Evolution of topics with respect to communities:}
Topics evolve with discussions.
New topics emerge, and existing topics spread out to reach more people.
Research, however, is yet to investigate the existence and nature of differences of discussions and evolution of topics, in different existing social communities, that start out with similar parameters (such as similar sized communities inside similar geography, with similar interest profiles, and at the same time) but show different adoption and evolution characteristics with respect to topics.

{\bf Evolution of topics with respect to geographies:}
Not much investigation has been carried out on how topics evolve differently across different geographies, and whether communities have a role to play in such evolution.
Further, the state-of-the-art research on, whether and how the discussions shape up differently in different geographies as topics evolve, is inadequate, and merits more focus.

\subsection{Social Structure Prediction}
\label{subsec:socstrucpred}
Literature exists for predicting links and social structures, using user-generated content, and topic-participation of users.
However, much work remains to be done.

{\bf Content based link prediction:}
Preliminary work has attempted to predict social links using user-generated content, such as in the work by \cite{puniyani2010social}.
However, deeper research is needed to identify and compare the effectiveness of different language models for content-based link prediction.
It will also be of interest to investigate whether such links act as high-conductance information paths, and be applied to situations demanding fast information diffusion.

{\bf Content based structure modeling:}
Literature needs to enhance the understanding of the social goodness of content-based links predicted, by analyzing and/or predicting implicit social structures that the links capture, such as social communities.
Being able to predict links that tend to form social structures accurately is a more rewarding solution than being able to predict stray links with the same accuracy, as the former would retain social properties (structures) of the social network.

\subsection{Impact of Social Structures}
\label{subsec:comm}
There is a dearth of literature that investigates social structures in combination with information diffusion, role of influence and topic modeling.
Some open challenges are mentioned below.

{\bf Impact of communities:}
No research study exists today, that attempts to investigate the impact of social network structures such as communities, on the generation and evolution of topics, on information diffusion or on the influence of individual users.
Yet, one would expect this to be of interest for researchers, as well as applications such as marketing and governance.

{\bf Aging of users in network and communities:}
There is a need of understanding whether and how the age of users in friendship-based and topical interest-based communities, as well as their age in the social network, impacts the information diffusion, the role of influence and topic dynamics.

\subsection{Role of Social Influence}
\label{subsec:roleofinfluence}
Many research works have studied the role of social influence of individual users in information diffusion and topics.
However, there is scope to improve in the following areas.

{\bf Impact of influential users on individuals:}
Some problems have remained unexplored in this area.
For instance, literature does not answer how information diffused on a social network will be accepted by a receiver, given receiver influence.
Do influential users also tend to diffuse more information compared to others?
Can one determine the set of seed users to inform initially, to maximize information reach?
In case the highly influential users do not tend to dissipate much information on a given topic (or in general), what is the combined impact of weak influencers?

{\bf Topic-specific influence:}
Existing topic and information spreading models, such the cascade and activation models, need to be adopted to incorporate influence.
The dynamics of users spreading and receiving influence with respect to topics requires research.
Also, whether influential users tend to create topic-specific communities and how that impacts information diffusion, needs investigation.

\subsection{Familiarity and Similarity}
\label{subsec:famsim}
Familiarity and similarity have not been studied much together, in the context of information diffusion, influence and topics taken together.
Homophily has been investigated in works such as \cite{weng2010twitterrank}.
However, there is a need to address familiarity and similarity together in intricate settings.

{\bf Consideration of social influence:}
Does social influence play a role in creating similarity among familiar people?
For example, do a pair of individuals become similar after becoming familiar?
Does one individual, among the pair, adapt become more like the other, and participate in similar discussions, generate similar topics, use similar language or belong to similar social communities?

{\bf Impact of communities and information diffusion:}
Literature does not explore the impact on similarity (topics, language usage {\it etc.}), given the evolution of friendships (familiarity), community participation and exposure to diffused information.
For instance, literature does not try to answer questions like, {\it If a person deeply connects to a community because of homophily, do their subsequent topics of interest and discussions align better with this community?}

{\bf Consideration of prior topics/interests:}
The existing studies related to familiarity and similarity do not consider the user profiles that capture the prior topics of user participations, or capture the explicit and latent interests of users.
This needs to be incorporated into the models.

\bibliographystyle{ACM-Reference-Format-Journals}
\bibliography{survey}

\end{document}